\input psfig
\tolerance=500
\vsize=8.4in
\voffset=0.24in
\hsize=6.0in
\hoffset=0.24in
\vbadness=10000
\baselineskip=14pt
\def\blankline{\par\vskip \baselineskip}
\def\refitem{\par\noindent\hangindent 20pt}
\newbox\grsign \setbox\grsign=\hbox{$>$} \newdimen\grdimen
\grdimen=\ht\grsign
\newbox\simlessbox \newbox\simgreatbox
\setbox\simgreatbox=\hbox{\raise.5ex\hbox{$>$}\llap
     {\lower.5ex\hbox{$\sim$}}}\ht1=\grdimen\dp1=0pt
\setbox\simlessbox=\hbox{\raise.5ex\hbox{$<$}\llap
     {\lower.5ex\hbox{$\sim$}}}\ht2=\grdimen\dp2=0pt
\def\simgreat{\mathrel{\copy\simgreatbox}}
\def\simless{\mathrel{\copy\simlessbox}}
\def\etal{{\it et~al{.}}}
\def\kms{\rm \ km\ s^{-1}}
\def\kmsmpc{\rm \ km\ s^{-1}\ Mpc^{-1}}

\font\noverm=cmr9
\font\novebf=cmbx9
\font\noveit=cmti9
\font\novesl=cmsl9
\font\novemit=cmmi9
\def\typenine{\let\rm=\noverm \let\bf=\novebf \let\it=\noveit
\let\sl=\novesl \let\mit=\novemit \rm}


\baselineskip=14pt
\blankline
\blankline

\centerline {\bf GALAXY PAIRWISE VELOCITY DISTRIBUTIONS }
\centerline {\bf ON NON-LINEAR SCALES}

\vskip 1.2in
\centerline {Antonaldo Diaferio and Margaret J. Geller}
\blankline
\blankline
\centerline {Harvard-Smithsonian Center for Astrophysics}
\centerline {60 Garden Street, Cambridge MA 02138}
\centerline {adiaferio@cfa.harvard.edu; mgeller@cfa.harvard.edu}

\blankline
\blankline
\centerline {{\it The Astrophysical Journal}, in press}
\vfill\eject
\blankline
\centerline{\bf ABSTRACT}
\blankline

\nobreak
The redshift-space correlation function $\xi_s$ for projected
galaxy separations
$\simless 1h^{-1}$ Mpc can be expressed as the convolution of the
real-space correlation function with the galaxy pairwise velocity
distribution function (PVDF).
An exponential PVDF yields the best fit to the $\xi_s$ measured
from galaxy samples of different redshift surveys.
We show that this exponential PVDF is not merely a fitting function but
arises from well defined gravitational processes.
Two ingredients conspire to yield a PVDF with a nearly
exponential shape: (i) the number density $n(\sigma)$ of systems with
velocity dispersion $\sigma$; (ii) the unrelaxed dynamical state of most
galaxy systems. The former ingredient determines the exponential tail and the
latter the central peak of the PVDF.

We examine a third issue: the transfer of orbital kinetic
energy to galaxy internal degrees of freedom. Although this effect
is of secondary importance for the PVDF exponential shape,
it is detectable in galaxy groups, indicating
that galaxy merging is an ongoing process in the present Universe.

We compare the $\xi_s$ measured on non-linear scales from galaxy samples
of the Center for Astrophysics redshift surveys
with different models of the PVDF convolved with the measured real-space
correlation function. This preliminary comparison indicates that the
agreement between model and observations depends strongly on both the
underlying cosmological model and the internal dynamics of galaxy systems.
Neither parameter dominates.
Moreover, the agreement depends sensitively on the
accuracy of the galaxy position and velocity measurements.

We expect that
$\xi_s$ will pose further constraints on the model of the Universe
and will improve the knowledge of the dynamics of galaxy systems on very
small scales
if we improve (i) the galaxy coordinate determination and (ii)
the measurement of relative velocities of galaxies with small projected
separation.
In fact, the redshift-space correlation function $\xi_s$ depends
sensitively on the internal pairwise
velocity distribution of individual galaxy systems for
projected pair separations $\simless 0.5h^{-1}$ Mpc and relative
velocities $\pi\simless 300\kms$.

\blankline \noindent
Subject Headings: Cosmology: Dark Matter -- Cosmology: Theory --
Galaxies: Clustering -- Galaxies: Interaction -- Gravitation

\vfill\eject
\blankline
{\bf 1. INTRODUCTION}

\nobreak
The pairwise velocity distribution function (PVDF) of galaxy systems has been
studied since Geller \& Peebles (1973) first used it to
determine the mean mass of galaxy groups statistically. The PVDF assumed
clear importance in cosmology when Peebles (1976) used it to determine
the pairwise velocity dispersion $\sigma_{12}(r)$ of galaxy pairs
separated by a projected distance $r\simless 1h^{-1}$
Mpc\footnote{$^1$}{$H_0=100h\kmsmpc$ is the present Hubble
constant and we use $h=0.5$ throughout.}
and ultimately to determine the mean mass density of the Universe.

Davis \& Peebles (1983) first computed the redshift-space
correlation function $\xi_s$ as a convolution of the real space
two-point correlation function with the galaxy PVDF.
Recently, Fisher {\etal} (1994b) and Marzke {\etal} (1995)
used the same ``convolution method'' to determine $\xi_s$ and the
pairwise velocity dispersion $\sigma_{12}(r)$ on non-linear scales.
All of this work, starting from Peebles (1976), assumes
an exponential PVDF. Bean {\etal} (1983), Fisher {\etal} (1994b) and
Marzke {\etal} (1995) demonstrate
quantitatively that this shape fits the observations better than other
distributions. However, so far the exponential shape has been a
fitting function without any physical justification.

It is not clear what PVDF we should expect on non-linear scale where
non-linear gravitational clustering erases information about the
initial conditions. On linear scales ($r\simgreat 10h^{-1}$ Mpc),
if the standard inflationary
model is valid, we expect a Gaussian PVDF (e.g. Nusser, Dekel, \& Yahil
1995). As a 2-point distribution, the PVDF is a more powerful tool than
1-point distributions to determine whether the density fluctuations
are Gaussian or non-Gaussian (e.g. Kofman {\etal} 1994;
Catelan \& Scherrer 1995).
If the PVDF is indeed Gaussian on linear scales, we need a link between
the observed exponential PVDF  on non-linear scales and the expected
Gaussian in the linear regime (Fisher 1995).

Numerical work devotes attention mainly to the pairwise velocity
dispersion $\sigma_{12}(r)$ rather than to the shape of the PVDF (e.g.
Couchman \& Carlberg 1992; Gelb \& Bertschinger 1994).
Efstathiou {\etal} (1988) simulate a flat Universe with scale free initial
conditions; by analyzing the particle velocity field,
they find a skewed PVDF with exponential tails but a flatter core at
small relative velocities. Cen \& Ostriker (1993) implicitly find the
same result by simulating a standard CDM Universe including dissipative
galaxy formation. Their single-galaxy peculiar velocity exponential
distribution implies a PVDF similar to the one found by Efstathiou {\etal}
(Marzke {\etal} 1995). A variety of CDM
models (Fisher {\etal} 1994b) confirm this behavior.
All of this work has a dynamic
range of roughly three orders of magnitude. With an order of magnitude
increase in dynamic range, Zurek {\etal}
(1994) study the massive halo velocity field and
find an exponential skewed PVDF at all projected separations
between $0.5h^{-1}$ Mpc and $5.5h^{-1}$ Mpc and for all relative
velocities, indicating that the flat core of previous simulations
probably arose from an inadequate treatment of gravitational
interactions on small scales.

All previous work, both observational and numerical, does not explain
the physical origin of the exponential shape of the PVDF.
Here we propose a simple physical argument for the observed exponential
PVDF  for galaxy separations $\simless 1h^{-1}$ Mpc.

If $n(\sigma)$ is
the number density of galaxy systems with velocity dispersion $\sigma$,
we show that the exponential tail can be obtained from
the integral  of Gaussian internal velocity distributions for
each galaxy system weighted either with the observed $n(\sigma)$
or with the $n(\sigma)$ predicted by the Press \& Schechter
(1974) theory (Sect. 2).

Sect. 3 shows that the central peak of
the PVDF requires the presence of unrelaxed systems with a non-Gaussian
internal velocity distribution. In Sect. 4 we examine a further process
which can peak up the PVDF at small relative velocities: the transfer of
orbital kinetic energy to galaxy internal degrees of freedom.
In Sect. 5 we compare various models of the PVDF with
the redshift-space correlation function measured
for the Center for Astrophysics (CfA) magnitude
limited redshift surveys.

\blankline
{\bf 2. THE PVDF FROM $n(\sigma)$}

\nobreak
Suppose that the probability of measuring one component
$u$ of the relative velocity  of two
galaxies within  a particular system is a universal function
$\Lambda(u,\sigma)$, where $\sigma$ is the velocity dispersion of the system
and $u$ is independent of the galaxy separation distance.
Assume that $n(\sigma)$ is the number density of systems with dispersion
$\sigma$. Moreover, assume that the number of galaxies $\nu$ within a system
with dispersion $\sigma$ depends only on $\sigma$: $\nu=\nu(\sigma)$.
The probability of choosing a single galaxy is $n(\sigma)
\nu(\sigma)$ and the probability of picking a galaxy pair within a
single system is $n^2(\sigma)\nu^2(\sigma)/n(\sigma)$.
Assume, for the sake of simplicity, that all the systems
are disjoint with separation $\simgreat 1h^{-1}$
Mpc. Thus the contribution to the pairwise velocity
distribution $p(u)$ for galaxy separation
$\simless 1 h^{-1}$ Mpc comes only from galaxy
pairs in the same system. Therefore, we can neglect the relative velocities
of systems. Then we have
$$ p(u,\sigma_{\rm min},\sigma_{\rm max})du
\propto du\int_{\sigma_{\rm min}}^{\sigma_{\rm max}}
\nu^2(\sigma) n(\sigma) \Lambda(u,\sigma)d\sigma. \eqno(2.1)$$

Hereafter, we refer to equation (2.1) as the PVDF.
Let us now make a few hypotheses which roughly approximate the internal
properties of galaxy systems. Let us assume that
systems have relaxed violently (Lynden-Bell 1967; Shu 1978).
We can then assume that systems approximate truncated singular
isothermal spheres with density profile $\rho(r)=\sigma^2/2\pi Gr^2$.
$N$-body simulations (Crone, Evrard, \& Richstone 1994; Carlberg 1994; Navarro,
Frenk, \& White 1996; Cole \& Lacey 1996) show
that this profile is not correct at very small and very large radii. However,
the slope $r^{-2}$ fits the dark halo density profile at least over
the range $0.1\simless r/r_{\rm vir} \simless 1$ where $r_{\rm vir}$ is
the radius containing an overdensity of 200 (Navarro {\etal} 1996).
We are interested in the relation between the galaxy number $\nu$ and
the velocity dispersion $\sigma$; thus the assumption
$\nu(\sigma)\propto\sigma^2$ is reasonable.
The isothermal model has a Gaussian velocity distribution. Thus
$$ \Lambda(u,\sigma)du = {1\over (4\sigma^2\pi)^{1/2}} \exp\left(-{u^2\over
4\sigma^2}\right)du. \eqno(2.2)$$

We have two choices for the number density $n(\sigma)$: (1) we can assume the
distribution derived from the Press \& Schechter (1974) theory which
approximates the number density of massive halos in $N$-body simulations of
a flat Universe with scale free initial conditions
(e.g. Efstathiou {\etal} 1988; Lacey \& Cole 1994);
(2) we can use the  observed distribution.

The Press-Schechter $n(\sigma)$ can be easily derived for
a flat CDM universe dominated by dissipationless dark matter.
 Following White \& Frenk (1991),
consider spherical perturbations with comoving radius $r_0$ which
have already collapsed into isothermal spheres by redshift $z$.
 For singular isothermal halos the velocity dispersion is
$\sigma^2 = GM(r)/2r$, independent of radius $r$.
If the halo mass is $M=4\pi\rho_0r_0^3/3$ where
$\rho_0$ is the present density of the Universe,
the velocity dispersion can be expressed in terms of the redshift and the
initial size of the perturbation: $\sigma = 1.68(1+z)^{1/2}
H_0r_0/\sqrt{2}$ where $H_0$ is the present Hubble constant
(however, see e.g. Jing \& Fang 1994; or Crone \& Geller 1995 for
mass-dispersion relations when the isothermal sphere approximation is
not valid).
The number of halos with dispersion $\sigma$ per comoving volume at
redshift $z$ is then
$$ n(\sigma)d\sigma = -{3(1.68)^3 H_0^3(1+z)^{3/2}\over
(4\pi)^{3/2} \sigma^4} {d\ln \Delta\over d\ln\sigma} \nu e^{-\nu^2/2}
d\sigma\eqno(2.3a)$$
where $\nu=\delta_c(1+z)/\Delta$ and $\delta_c=1.69$ is the mean linear
interior mass overdensity
when each spherical shell recollapses to the origin (Narayan \& White
1988). The rms linear mass overdensity in a sphere of radius $r_0$ is
$$ \Delta(r_0) = 16.3\sigma_8
(1-0.3909r_0^{0.1}+0.4814r_0^{0.2})^{-10}\eqno(2.3b)$$
where $\sigma_8^2$ is the usual ratio of the variances of the mass and galaxy
fluctuations within randomly placed spheres of radius $8h^{-1}$ Mpc.
Equation (2.3b) approximates  the correct $\Delta(r_0)$
to within 10\% over the range $0.03h^{-1} {\rm Mpc}
<r_0<20h^{-1} {\rm Mpc}$. The correct $\Delta(r_0)$ is obtained through the
convolution of the CDM linear power spectrum
with the spherical top-hat window function of radius $r_0$.
The power spectrum, assuming $\Omega_0=1$, $h=0.5$ and a cosmic
microwave background temperature $\theta = 2.7$ $^o$K, is (Davis {\etal} 1985)
$$ P(k) = 1.94\times 10^4 \sigma_8^2 {k\over (1+6.8k +72k^{3/2} +
16k^2)^2 } {\rm Mpc}^3. \eqno(2.3c) $$
The only free parameter is now the normalization parameter $\sigma_8$.

A sample of 25 Abell clusters with velocity dispersion $\sigma >
300\kms$ and 31 galaxy groups in the CfA
redshift surveys with $\sigma>100\kms$ (Zabludoff {\etal} 1993b) yields
$$ n(\sigma)d\sigma \propto 10^{-\alpha\sigma}d\sigma \eqno(2.4)$$
where $\alpha=0.0015$. Equation (2.4) holds for $\sigma>700\kms$. Mazure
{\etal} (1996) analyze a volume-limited sample of 128 Abell clusters
with richness $R\ge 1$. They find a similar $n(\sigma)$
with $\alpha\sim 0.0016$ for $\sigma>800\kms$. In both cases, the
distribution is shallower for smaller $\sigma$. Thus, using equation (2.4)
for the whole range of $\sigma$ overestimates the number of system with
small velocity dispersion. However, such an overestimate does not affect
our analysis. In fact, we shall see that we need even more
systems with $\sigma<700\kms$ to obtain an exponential PVDF.

With these assumptions the integral in equation
(2.1) at $z=0$ yields the PVDFs in
Fig. 1 for different values of $\sigma_{\rm min}$ and for $\sigma_8=0.5$,
$1.0$, and $1.5$, to map the range of COBE normalizations for different
CDM models (Bunn, Scott \& White 1995). We set $\sigma_{\rm max}=1500\kms$.

If we decrease $\sigma_{\rm min}$, the Press-Schechter $n(\sigma)$ includes
a larger fraction of halos of small size and peaks up the center of the
distribution. However, most halos with $\sigma\simless 150 \kms$ are
likely to  contain at most one galaxy as luminous as the Milky Way.
Systems of galaxies with $\sigma\simless 100\kms $ are only a small fraction of
the total number of systems predicted by the Press-Schechter theory at these
velocity dispersions. Moreover, it is well known that
the Press-Schechter $n(\sigma)$ for a flat CDM universe
overestimates the observed number of single galaxy halos with
$\sigma\simless 100\kms$ (see e.g. White 1993). If we set
$\sigma_{\rm min}\sim 100 \kms$ as the lower limit of
integration in equation (2.1), we mostly exclude halos containing only a
single galaxy.

The observed $n(\sigma)$ does not include individual galaxies by definition
and does not overestimate the number of galaxy systems with small
$\sigma$. Therefore, the PVDF does not change appreciably
for any choice of $\sigma_{\rm min}$.

\topinsert
\baselineskip=10pt
\moveleft 0.1\hsize
\vbox to .8\vsize {\hfil
\psfig{figure=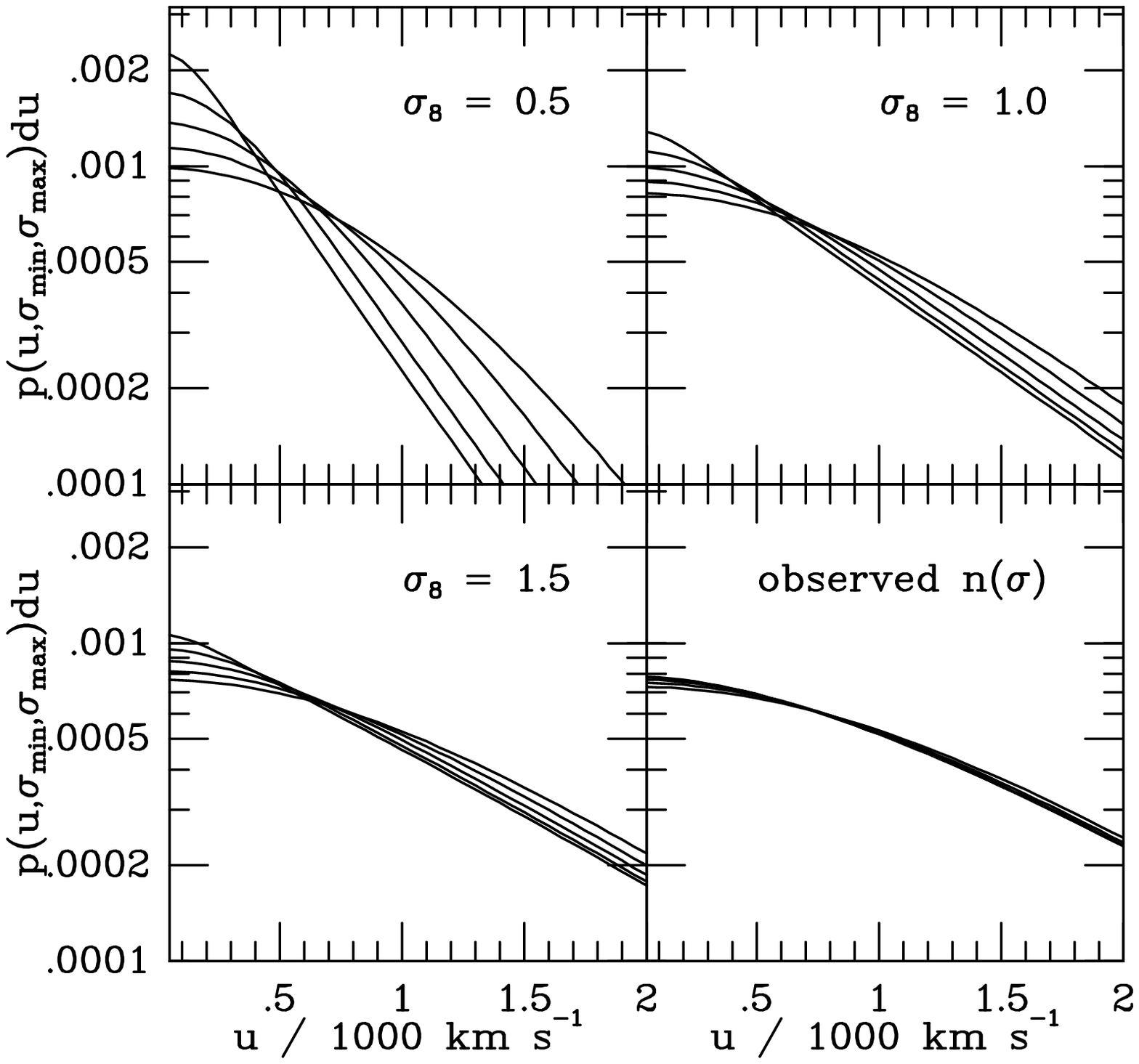,height=.6\vsize,width=0.8\hsize}\hfil
}
\vskip -0.4\vsize
\noindent
{\typenine {\bf Figure 1}
The PVDF computed through equation (2.1) with
a Gaussian internal pairwise velocity distribution $\Lambda(u,\sigma)$.
The number density of galaxy systems is
the observed $n(\sigma)$ (eq. [2.4]) or
the Press-Schechter $n(\sigma)$ for a flat CDM universe
(eqs. [2.3]) with different normalization
$\sigma_8$. We take $\sigma_{\rm max} = 1500$ km s$^{-1}$. At large relative
velocities $u$, from top to bottom, the
curves in each panel have $\sigma_{\rm min} = 500$, 400, 300, 200, and
100 km s$^{-1}$, respectively.  }
\endinsert

Fig. 1 shows that for reasonable values of $\sigma_{\rm min}\simgreat
100 \kms$, equation (2.1)
predicts a PVDF which is almost exponential at large relative velocities
$u$.  However, the PVDF bends over at smaller $u$.
In order to obtain an exponential core
we need a different model for $\Lambda(u,\sigma)$.

\blankline
{\bf 3. $\Lambda(u,\sigma)$ OF UNRELAXED SYSTEMS}

\nobreak
It is well known that steady-state self-gravitating systems cannot
have exactly Gaussian velocity distributions because
escaping stars deplete the high-velocity  tails (e.g. King 1965, 1966).
In fact, the isothermal sphere is the only self-gravitating system with a
Gaussian velocity distribution. However, its mass is infinite and real steady
state systems tend only asymptotically to the Gaussian
distribution (e.g. Padmanabhan 1990).

We should not expect a Gaussian distribution in galaxy systems for
another reason: many observed galaxy systems -- from groups to clusters --
are unlikely to be relaxed. Most galaxy groups are still collapsing
(e.g. Diaferio {\etal} 1993; Doe {\etal} 1995)
 and many clusters contain substructures which indicate
that they are far from equilibrium (e.g. West, Jones, \& Forman 1995;
Colless \& Dunn 1996). Therefore,  a single Gaussian is not a good
approximation to their velocity distribution.
Velocity distributions will depend on the initial conditions and on the
dynamical state of the system. In general, there will not be a
universal $\Lambda(u,\sigma)$ for all galaxy systems.

These arguments apparently show that the assumptions about the shape and
the uniqueness of the distribution $\Lambda(u)$ in Sect. 2 are
inadequate. However, we
can investigate what shape $\Lambda(u)$ tends to assume
in a hierarchical clustering scenario, where either systems are
virialized or they are still forming through the aggregation of
virialized subunits.

To model the evolution of a cluster by accretion of subunits,
the excursion set formalism (see e.g. Bower 1991; Bond {\etal} 1991;
Kauffmann \& White 1993; Lacey \& Cole 1993)
extends the Press-Schechter formalism to estimate
the number density of halos with mass $M_1$ at redshift
$z_1$ which will merge at different times to form a single halo
of mass $M_2>M_1$ at redshift $z_2<z_1$. We have
$$ \eqalignno{ n(M_1,z_1\vert M_2,z_2) dM_1 =&  \left(2\over \pi\right)^{1/2}
{\rho_0\over M_1} \left\vert{d\ln\Delta_1\over dM_1}\right\vert
{ (\delta_1 - \delta_2) \Delta_1^2 \over
(\Delta_1^2 -\Delta_2^2)^{3/2}}\times \cr
 & \times \exp\left[-{(\delta_1-\delta_2)^2\over
2(\Delta_1^2 -\Delta_2^2)}\right] dM_1 &  (3.1a)\cr} $$
where $\rho_0$ is the present density of the Universe, $\Delta_i^2$ ($i=1$,
2) is the variance of the linear overdensity in a sphere containing the mass
$M_i$, and $\delta_i=1.69/D(z_i)$ is the
extrapolated linear critical overdensity
for which perturbations with overdensity $\delta>\delta_i$ have collapsed
at redshift $z_i$. $D(z_i)$ is the perturbation growth factor.

Following the procedure used to obtain the number density of halos with
velocity dispersion $\sigma$ (eq. [2.3a]),
we can use equation (3.1a) to express the number density of halos with
velocity dispersion $\sigma_1$ at redshift $z_1$ which will form a halo
with dispersion $\sigma_2$ at redshift $z_2<z_1$. In a flat CDM universe,
$D(z_i)=(1+z_i)^{-1}$. Mass and dispersion of each halo are related by
$M_i=4\pi\rho_0[r_0^{(i)}]^3/3$ and
$\sigma_i=1.68(1+z_i)^{1/2}H_0r_0^{(i)}/\sqrt{2}$. Equation (3.1a) becomes
$$ n(\sigma_1,z_1\vert\sigma_2,z_2) d\sigma_1 =
- {3(1.68)^3H_0^3(1+z_1)^{3/2} \over (4\pi)^{3/2} \sigma_1^4}
{d\ln\Delta_1\over d\ln \sigma_1} {\Delta_1^2\over
\Delta_1^2-\Delta_2^2} \tilde\nu e^{-\tilde\nu^2/2} d\sigma_1\eqno (3.1b)$$
where $\tilde\nu = 1.69(z_1-z_2)/(\Delta_1^2-\Delta_2^2)^{1/2}$ and we
express $\Delta_i$ with the approximation of equation (2.3b).

Let us now suppose that we observe a galaxy system at
redshift $z_1=z$ which has not yet collapsed, but it still contains different
substructures which will merge to form a single halo
with velocity dispersion $\sigma_2=\sigma_{\rm max}$
at a later epoch, e.g. $z_2= 0$.
We want to compute the probability $\Lambda(u)$ of measuring a velocity
difference $u$ between two galaxies within this collapsing system
at redshift $z$.

\topinsert
\baselineskip=10pt
\moveleft 0.1\hsize
\vbox to .8\vsize {\hfil
\psfig{figure=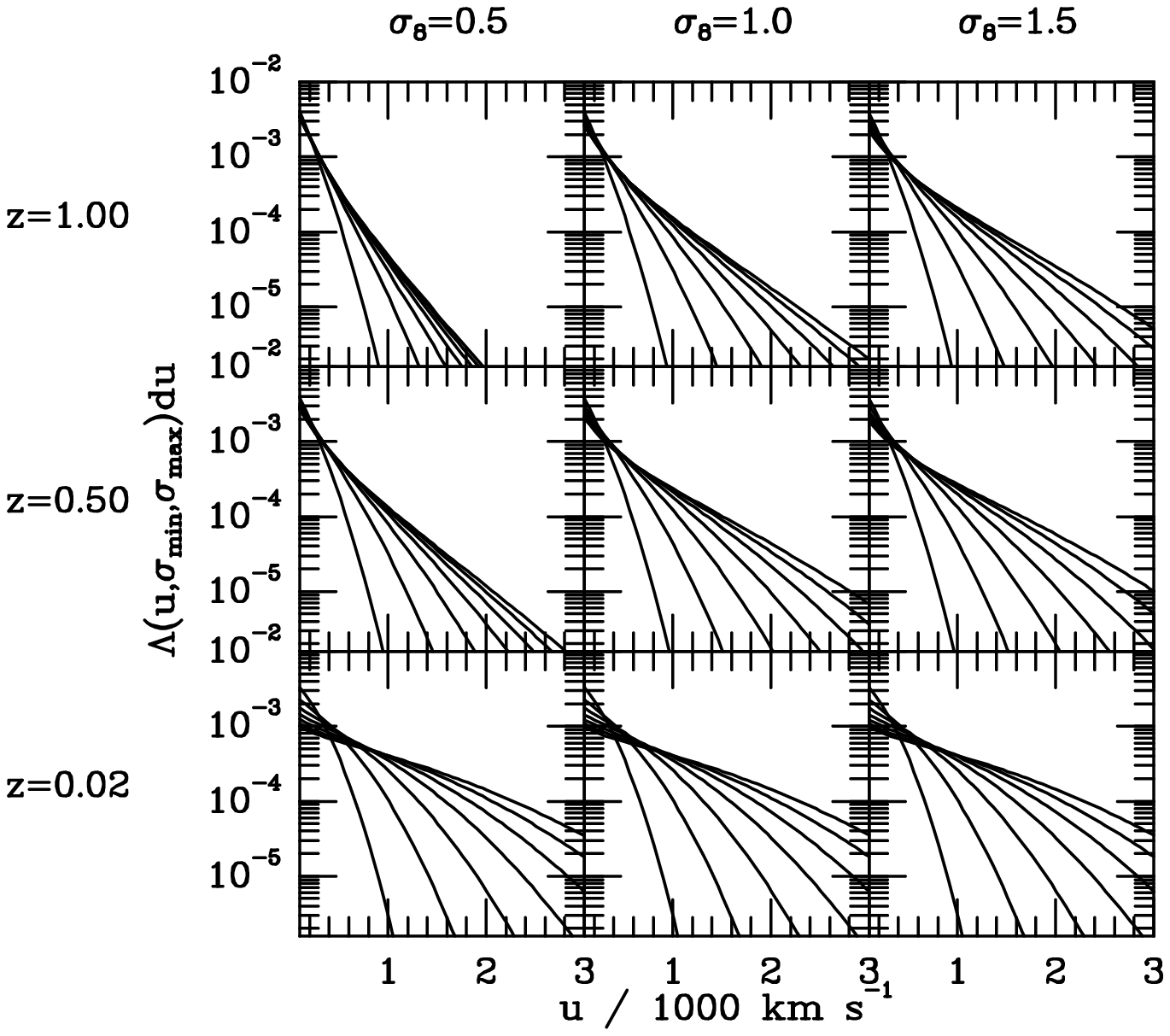,height=.6\vsize,width=0.8\hsize}\hfil
}
\vskip -0.5\vsize
\noindent
{\typenine {\bf Figure 2} Pairwise velocity difference distribution
$\Lambda(u)$ for a single unrelaxed galaxy system (eq. [3.2b]) at
different redshift $z$ and with different normalization $\sigma_8$.
We take $\sigma_{\rm min} = 100$ km s$^{-1}$. At large relative velocities
$u$, from top to bottom, the curves in each panel have $\sigma_{\rm
max} = 1500$, 1300, 1100, 900, 700, 500, and 300 km s$^{-1}$, respectively.  }
\endinsert

If we assume that each subunit has a
velocity distribution $\psi(v,\sigma)$, the probability of choosing a
galaxy with velocity $v$ within the system is
$$\alpha(v,\sigma_{\rm min},\sigma_{\rm max},z)dv \propto dv
\int_{\sigma_{\rm min}}^{\sigma_{\rm max}}
n(\sigma,z\vert\sigma_{\rm max},0)\nu(\sigma)\psi(v,\sigma)d\sigma
\eqno (3.2a)  $$
where $n(\sigma,z\vert\sigma_{\rm
max},0)$ is given by equation (3.1b) and $\nu(\sigma)$ is the number
of galaxies within each substructure as in Sect. 2. For the sake of
simplicity, equation (3.2a) ignores
the relative velocities of the subunits. The inclusion of this effect
will probably broaden the
velocity distribution $\alpha$. Thus, equation (3.2a) is conservative with
respect to our purpose of investigating the departure of $\alpha$ from a
Gaussian. Here however, we limit our analysis to the simplest case.

We notice that equation (3.2a) is  a velocity distribution decomposition
if we look at it the other way around. In other words, we ``build'' the
velocity distribution instead of decomposing it. For example,
van der Marel \& Franx (1993) decompose line profiles of elliptical
galaxies in orthogonal Gauss-Hermite functions to quantify departures from
Gaussian line profiles. Zabludoff, Franx, \& Geller (1993a)
apply a discrete version of this technique to the velocity
distributions of eight rich Abell clusters.
Here, we point out that the hierarchical clustering scenario
naturally predicts that $\alpha(v)$ is a sum of elementary distributions.

Equation (3.2a) quantifies the degree of subclustering in a galaxy
system. Comparison of equation (3.2a) with velocity
distributions of real clusters may provide constraints on the density of
the Universe (e.g. Evrard {\etal} 1993, Jing {\etal} 1995).
Moreover, extensions of equation (3.2a) can determine the rate of growth
of clusters as a function of redshift (Lacey \& Cole 1993).

Now we can write the pairwise velocity distribution $\Lambda(u)$ as
$$\eqalignno{\Lambda(u,\sigma_{\rm min},\sigma_{\rm max},z)du \propto &
du \int \alpha(v_1)\alpha(v_2)
\delta(\vert v_1-v_2\vert -u) dv_1dv_2 \cr
 = & du \int \alpha(v_1)\alpha(v_1+u) dv_1. & (3.2b) \cr} $$

Let us assume that the subunits
are virialized and that each approximates an isothermal sphere.
Therefore, $\psi(v,\sigma)$ is Gaussian and $\nu(\sigma) \propto
\sigma^2$ as in Sect. 2.  Integration of equation (3.2b) yields the
curves in Fig. 2 for different values of $\sigma_{\rm max}$, i.e. for
different masses of the final dark halo, and for
$\sigma_{\rm min}=100\kms$. Fig. 2 shows that as we decrease the redshift
$z$, i.e. as we come closer to the formation of the final system,
$\Lambda(u)$ approaches a Gaussian, as expected. At high redshift the
system is far from equilibrium and $\Lambda(u)$ is more centrally
peaked.

Is the presence of substructure the only physical
process responsible for more centrally peaked $\Lambda(u)$'s? Formally, the
presence of substructures implies that $\Lambda(u)$ is a
weighted integral of elementary distributions
(eqs. [3.2]). This assumption is common to other fields: weighted
integrals of Gaussian distributions  are also invoked to explain
the exponential shape of molecular cloud emission lines (e.g. Ida \& Taguchi
1996; but see also Miesch \& Scalo 1995) or the small scale velocity
gradient distribution in turbulent flows (Castaing, Gagne, \& Hopfinger 1990;
see also She 1991).

However, there is a very simple example in gravitational dynamics where a
centrally peaked $\Lambda(u)$ does not originate from a weighted integral.
Consider a collapsed region subject to secondary infall (Gunn \& Gott
1972): infalling galaxies
have small relative velocities and peak up the center of the
distribution whereas galaxies in the virialized region populate the
exponential tails.

An $N$-body toy model illustrates this issue. Consider
an isolated mass sphere with initial radial density $\rho(r) = (M/4\pi
R^3) (R/r)^2$, where $R$ is the radius and $M$ the total mass of the
sphere. If each shell is initially expanding according to the Hubble
flow $\dot r(0) = H_0r_0$, the maximum expansion radius is $r_{\rm max}
= r_0 \exp(B)$ at time $t_{\rm max} = H_0^{-1}
\sqrt{\pi B} \exp(B) P(1/2, B)$, where $B=H_0^2r_0^2R/2GM$ and
$P(\alpha,x)$ is the incomplete gamma function.

We choose this density profile for two reasons: (1) the difference
between the time and the radius at maximum expansion of two different shells
grows exponentially  and distinguishes the infall and the
virialized regions clearly for our illustrative purpose; (2) the virialized
region has the quasi-equilibrium density profile
$r^{-2}$. Thus, at the same time, we minimize the relaxation process and
isolate the effect due to the infall region.

\topinsert
\baselineskip=10pt
\moveleft 0.1\hsize
\vbox to .9\vsize {\hfil
\psfig{figure=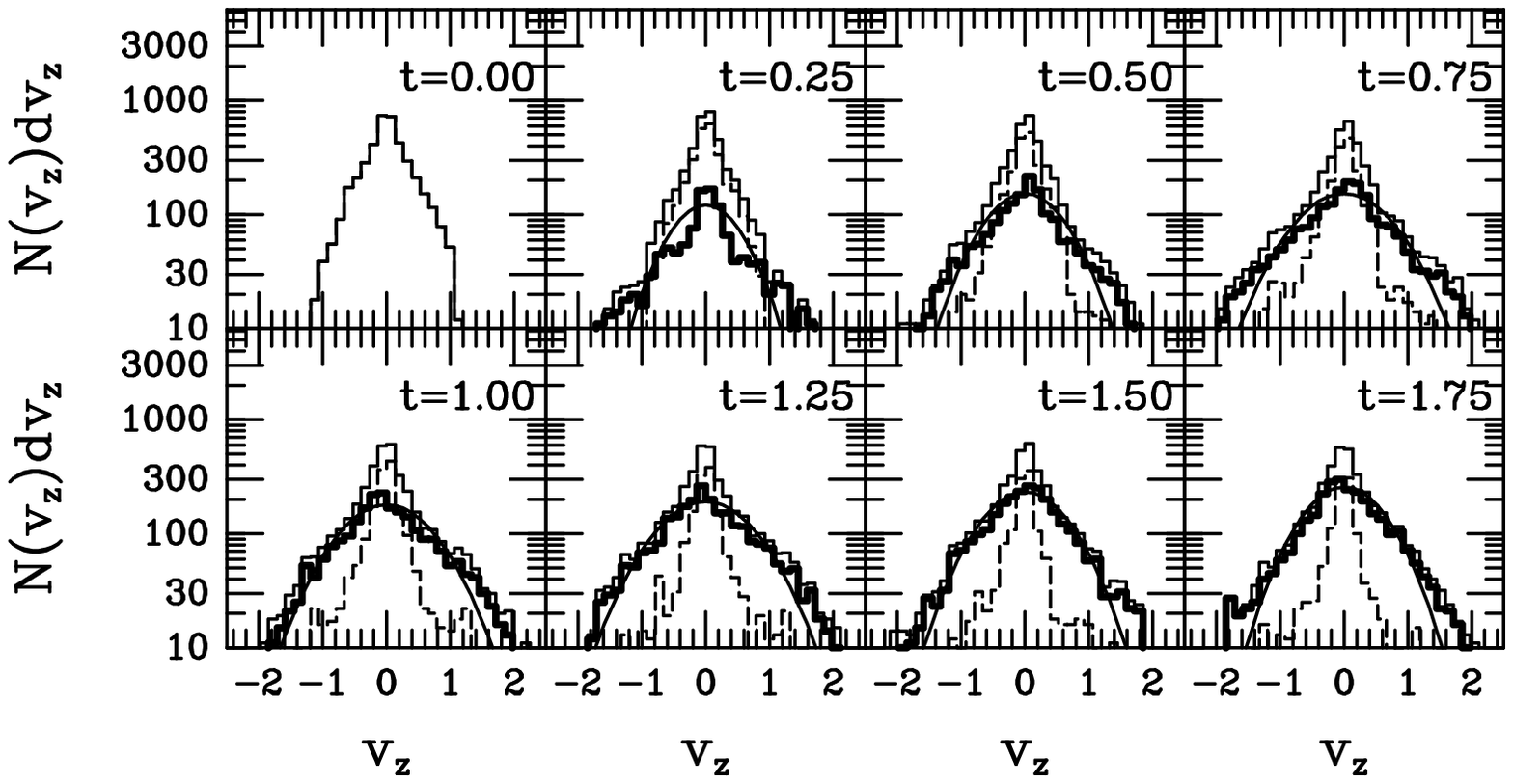,height=.7\vsize,width=0.8\hsize}\hfil
}
\vskip -0.63\vsize
\noindent
{\typenine {\bf Figure 3a}
Evolution of the distribution of the velocity
components $v_z$ for an isolated spherical system of particles
initially expanding with
the Hubble flow and with (a) an initial $r^{-2}$ density profile
or (c) an initial $r^{-1}$ density profile.
Times are in units of the collapse time $t_c$. The solid
histogram is the total distribution, the bold histogram is the
distribution for particles within the virialized core and the
dashed histogram for particles within the infall region. The curves are
the best Gaussian fits to the virialized core distributions.
Panels (b) and (d) show the evolution of the pairwise velocity
difference distribution $\Lambda(u)$ for the system with
initial $r^{-2}$ or $r^{-1}$ density profiles, respectively.}


\baselineskip=10pt
\moveleft 0.1\hsize
\vbox to .9\vsize {\hfil
\psfig{figure=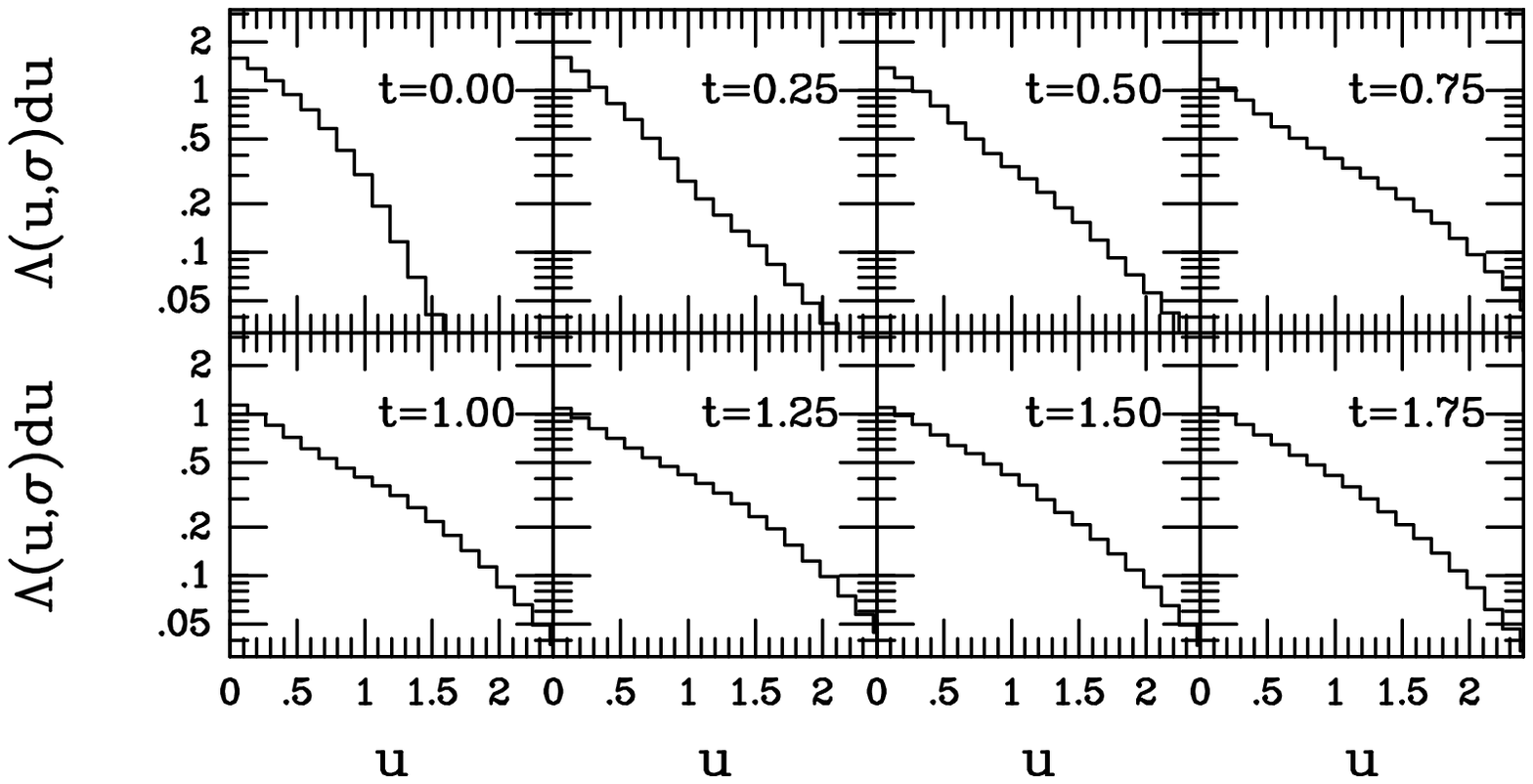,height=.7\vsize,width=0.8\hsize}\hfil
}
\vskip -0.63\vsize
\noindent
{\typenine \centerline{\bf Figure 3b}}
\endinsert

We follow the evolution of an isolated spherical system with $N=4096$
particles
initially expanding with the Hubble flow. In numerical units, the
gravitational constant is $G=1$, the system has total mass $M=1$ and
radius
$R=1$, and the initial Hubble constant is $H_0=1.2$.
We use the TREECODE by Hernquist (1987)
with softening parameter $\varepsilon=0.02R$ and tolerance parameter
$\theta=0.8$. We integrate the particle equations of motion
for two collapse times
$t_c=2\pi(3/10)^{3/2}GM^{5/2}/\vert E\vert^{3/2}$, where $E$ is the
total energy of the system.
The integration time step is $\Delta t=10^{-3}t_c$.

Fig. 3a shows the evolution of the distribution of the
velocity component $v_z$. We
show three distributions at each time: the total distribution (solid
histogram), the virialized region distribution (bold histogram) and
the infall region distribution (dashed histogram). At each time $t$,
particles with $4t_{\rm max}\le t$ belong to the virialized region and
particles with $4t_{\rm max}>t$ belong to the infall region. We
superimpose the best Gaussian fit on the virialized region distribution
to show that virialization indeed took place in the central region. We
choose the time limit $4t_{\rm max}>2t_{\rm max}$ to
suppress oscillation effects. Fig. 3a clearly shows that the infall
region is responsible for the central peak and the virialized region is
responsible for the tails of the total distribution.
Fig. 3b show that the total distributions
in Fig. 3a yield nearly exponential $\Lambda(u)$'s.

We also ran a simulation similar to the one above but with an initial
density profile $\rho(r)=(M/2\pi R^3)(R/r)$ which yields
$r_{\rm max} = H_0^2r_0^2/2B + r_0$ and $t_{\rm max}=H_0r_0/B$ where
$B=GM/R^2$. The system relaxes  faster than with an initial  $r^{-2}$ density
profile and secondary infall lasts for only a small fraction of the collapse
time (Fig. 3c).
However, when secondary infall involves a large mass fraction of the
sphere, $\Lambda(u)$ is exponential (Fig. 3d).

Velocity distributions like those in Fig. 3a are difficult to observe
in individual real systems. Infall regions where galaxies have small
velocities relative to the center of mass of the cluster are close to
the
turnaround radius which is $\simgreat 1h^{-1}$ Mpc for a typical Abell
cluster with mass $\sim 10^{14} h^{-1}$ M$_\odot$, i.e. richness $R=1$.
Those regions are contaminated
by foreground and background objects and they are generally poorly
sampled.
Therefore, when systems are observed individually,
observational biases imply that departures from Gaussian velocity
distributions in observed  clusters are more likely
due to subclustering rather than to infall region effects.
However, both effects are present in redshift surveys where large
regions
of the Universe are sampled.

\topinsert
\baselineskip=10pt
\moveleft 0.1\hsize
\vbox to .9\vsize {\hfil
\psfig{figure=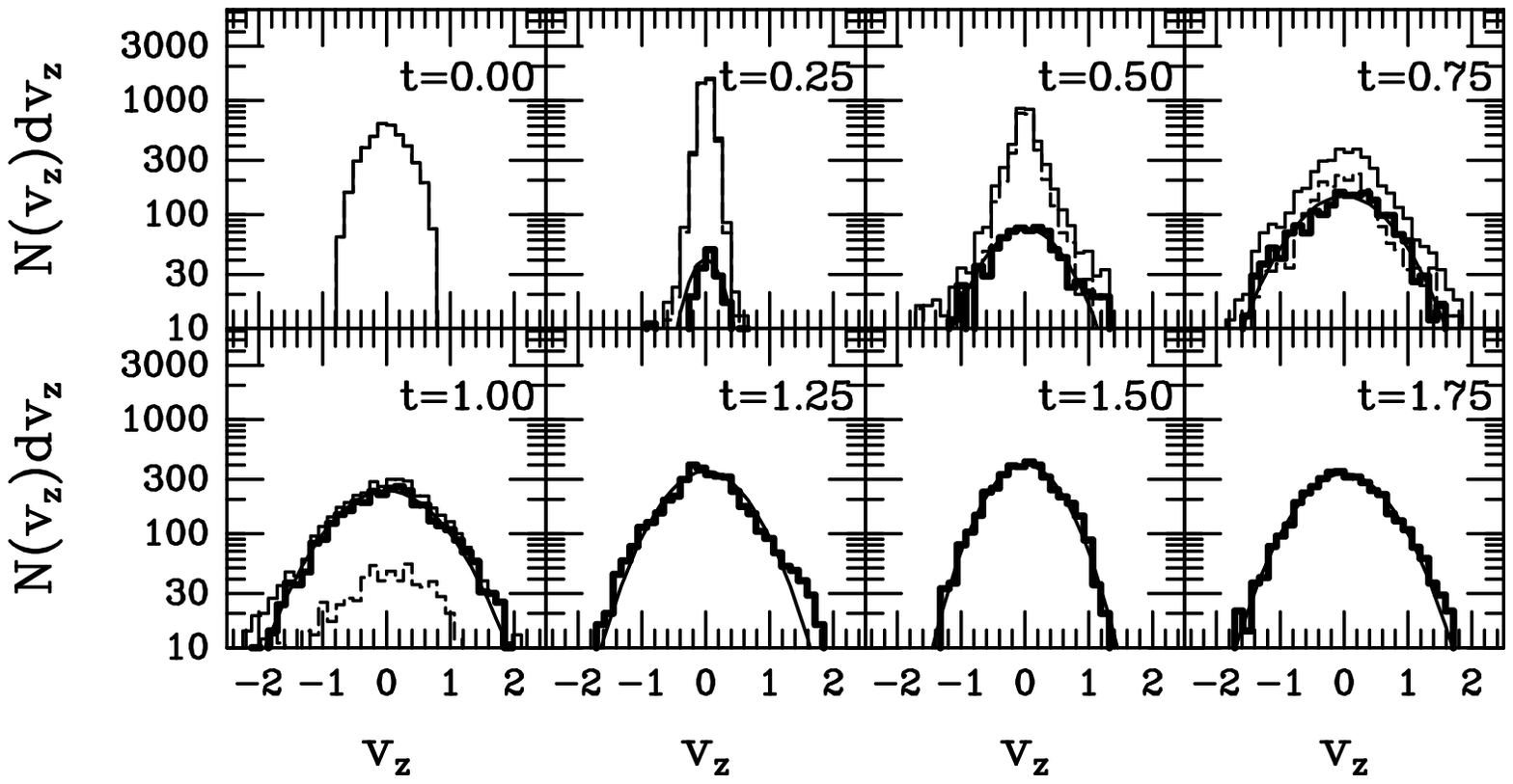,height=.7\vsize,width=0.8\hsize}\hfil
}
\vskip -0.63\vsize
\noindent
{\typenine \centerline{\bf Figure 3c}}

\baselineskip=10pt
\moveleft 0.1\hsize
\vbox to .9\vsize {\hfil
\psfig{figure=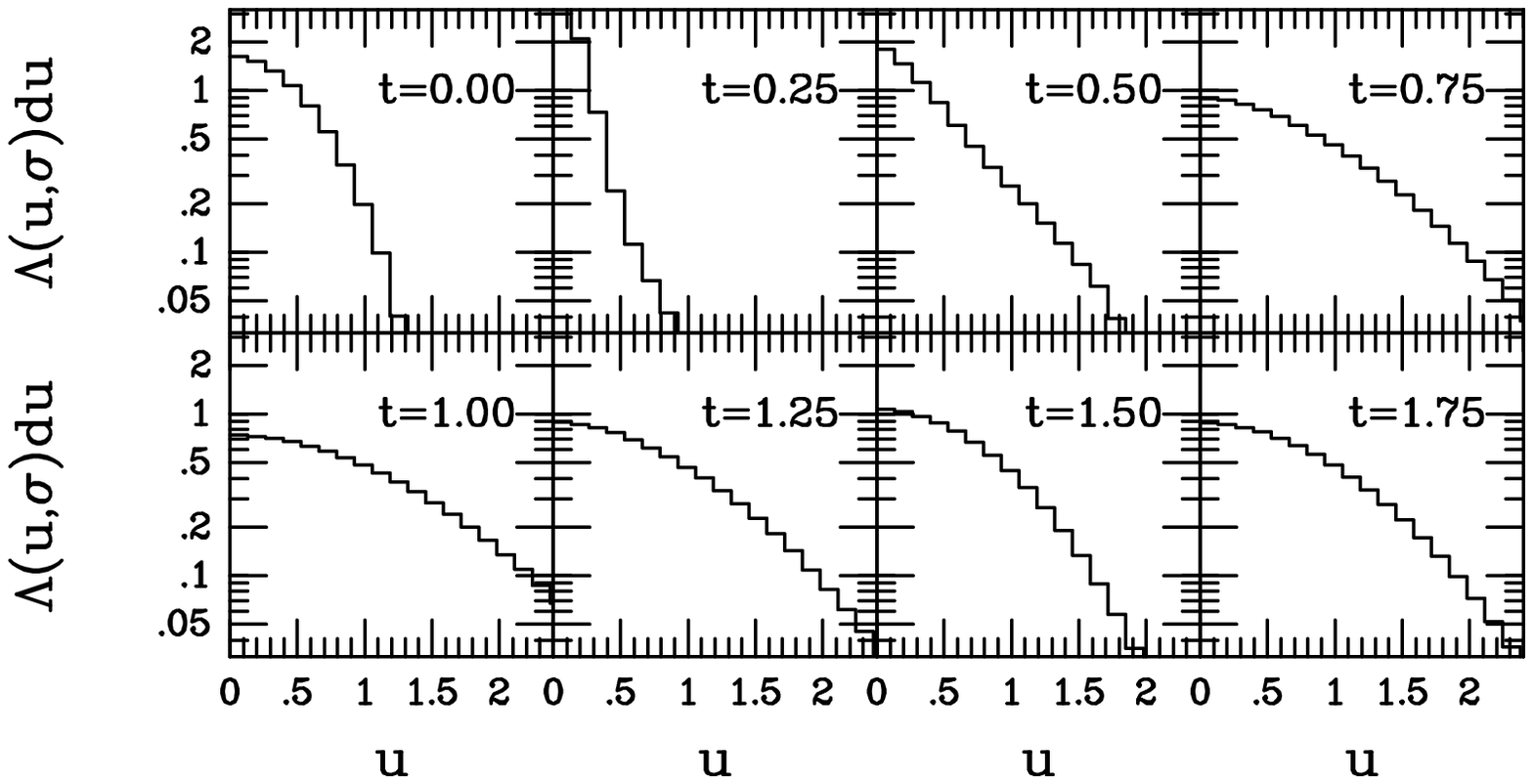,height=.7\vsize,width=0.8\hsize}\hfil
}
\vskip -0.63\vsize
\noindent
{\typenine \centerline{\bf Figure 3d}}
\endinsert

The preceding discussion shows that unrelaxed systems imply
that $\Lambda(u)$'s differ from Gaussians.
The $\Lambda(u)$'s for unrelaxed systems tend to have a more pronounced
central peak than  Gaussian distributions.
A combination of substructures and infall regions
contribute to this shape.

It is now clear that a unique and universal
$\Lambda(u,\sigma)$ does not exist, but rather each system has a
distribution depending on its particular dynamical state.
However, for the sake of investigation,
we persist in the assumption of a universal $\Lambda(u)$
and we examine the way the PVDF $p(u)$ in equation (2.1)
changes when $\Lambda(u,\sigma)$ differs from a
Gaussian. Figs. 2 and 3 suggest that systems far from equilibrium may have a
distribution similar to an exponential
$$\Lambda(u,\sigma)du = {1\over \sigma\sqrt{2}} \exp
\left(-\sqrt{2}\vert u\vert\over \sigma\right)du. \eqno(3.3)$$

\topinsert
\baselineskip=10pt
\moveleft 0.1\hsize
\vbox to .8\vsize {\hfil
\psfig{figure=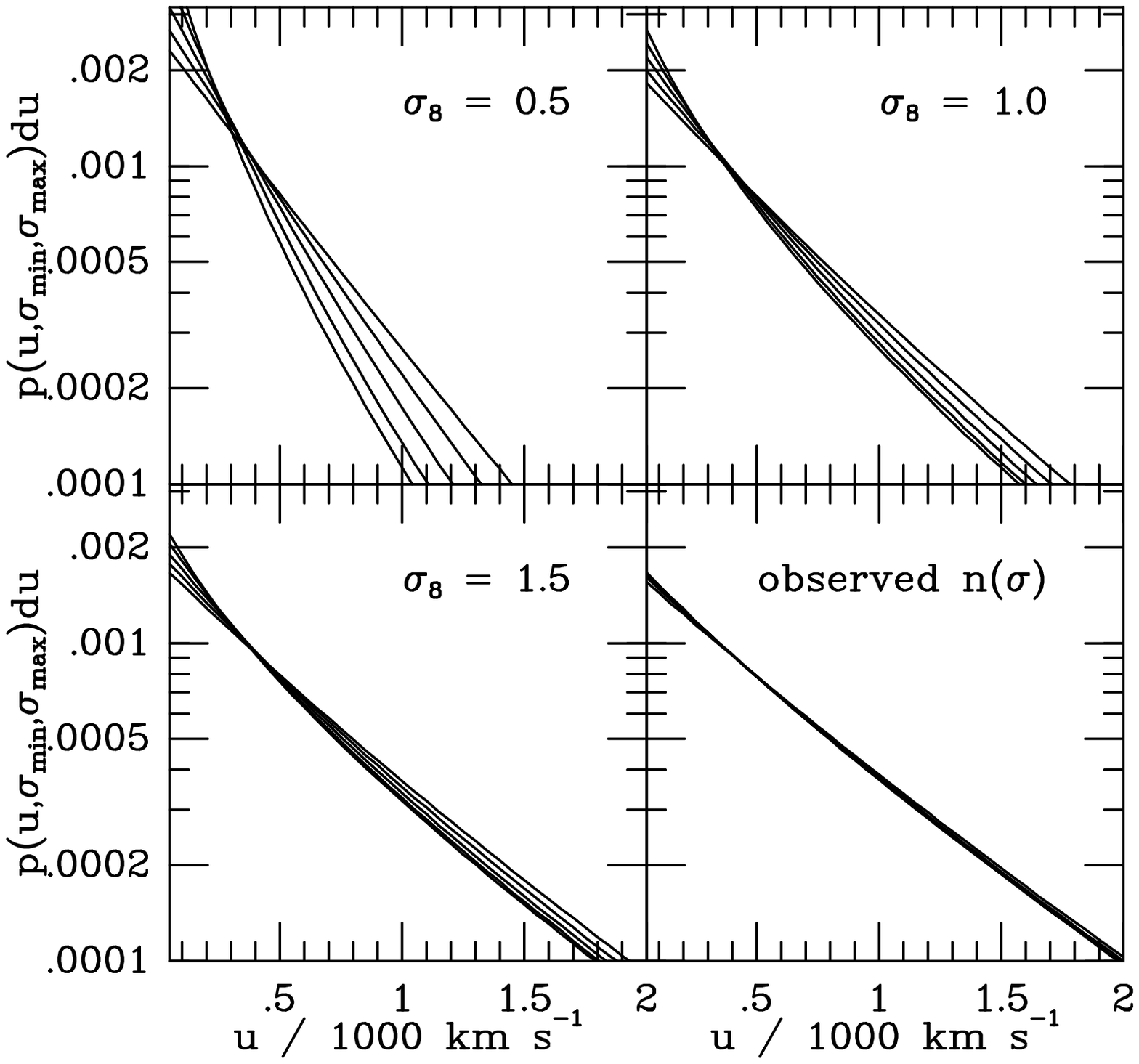,height=.6\vsize,width=0.8\hsize}\hfil
}
\vskip -0.4\vsize
\noindent
{\typenine {\bf Figure 4} Same as Fig. 1, but with an exponential
$\Lambda(u,\sigma)$ (eq. [3.3]).}
\endinsert

With an exponential $\Lambda(u)$ the integral in equation
(2.1) yields the curves in
Fig. 4. We clearly see that the central peak is more pronounced than
in the Gaussian case (Fig. 1) yielding a better approximation to
an ``exponential'' shape for the PVDF.

We stress here that the results of Figs. 1 and
4 are based on the generic assumption that
all galaxy systems have the same internal pairwise velocity
distribution $\Lambda(u,\sigma)$. However, the use of a unique
$\Lambda(u,\sigma)$ is still meaningful if we interpret
$\Lambda(u,\sigma)$ as a convolution of different internal pairwise
velocity distributions of individual systems at different dynamical
states but with the same $\sigma$. Thus, the results of Figs.
1 and 4 allow us to
conclude that if the real PVDF is indeed exponential over a wide range
of relative velocities $u$, the blend of different internal pairwise
velocity distributions
must contain a large fraction of distributions more centrally peaked
than a Gaussian. In other words, the exponential shape of the PVDF is a
signature of the presence of a large fraction of unrelaxed
systems.

\blankline
{\bf 4. $\Lambda(u,\sigma)$ OF DISSIPATIVE SYSTEMS}

\nobreak
In Sect. 3 we show that unrelaxed  systems have
internal pairwise velocity distribution $\Lambda(u,\sigma)$ more
centrally peaked than a Gaussian. This shape arises from
secondary infall and the presence of substructures.
Here we investigate a third physical process which can peak up
the $\Lambda(u)$ distribution at small relative
velocities: the transfer of orbital
kinetic energy to galaxy internal degrees of freedom.
In fact, this effect has only a secondary impact on the velocity
distribution.

To examine this issue, one should solve the complete Boltzmann equation for the
evolution of the phase space density of the galaxy system including
a collisional term. Fusco-Femiano \& Menci (1995) study how the velocity
distribution evolves in the presence of binary mergers in an external
gravitational potential.
In other words, they compute the decrease of orbital
kinetic energy when binaries disappear by merging. Here, we want
to include the orbital kinetic energy loss due to tidal
perturbations, thus computing the energy loss when  mergings do not take place.

We assume a simple physical model to derive an analytic
expression for the expected $\Lambda(u)$.
Assume an initially stable self-gravitating gas of particles.
Imagine switching on the particle internal degrees of freedom at a
time $t_0$. Now, during the motion of
particles within the system, tidal effects increase
the particle internal energy
at the expense of the orbital kinetic energy of the particles.
This system is apparent unstable, tending ultimately to a general
merging of `hot' particles. This process indeed occurs in galaxy
groups (see e.g. Mamon 1992b; Diaferio {\etal} 1993; Doe {\etal} 1995;
Weil \& Hernquist 1996).

The most serious shortcoming of this model is that we assume
a constant particle mass, clearly incorrect because merging
and tidal stripping are ongoing processes. Both processes increase the
kinetic energy loss. When galaxies merge, a binary system disappears and its
relative kinetic energy is completely transferred to the internal energy
of the remnant. Tidally stripped matter
forms a common background envelope. Particle
cores also lose kinetic energy through dynamical friction against this
background. Therefore, the main consequence of ignoring mass loss is to
underestimate the kinetic energy loss.

However, with these hypotheses and the assumption that the unperturbed
velocity distribution is Gaussian, we derive
the perturbed distribution (see the Appendix)
$$ \Lambda(u,\sigma)du = \tilde C(\sigma,\alpha,\varphi) \exp\left(-{u^2\over
4\sigma^2}\right) \left[1-\alpha\tilde H\left({u\over\sigma \sqrt{2}},
\varphi\right)\right]du \eqno(4.1a) $$
where $\tilde C$ is the normalization constant and
the function $\tilde H$  can be expressed in terms of the modified
Bessel functions.  In addition to the velocity dispersion $\sigma$,
this distribution depends on two
parameters $\alpha$ and $\varphi$. If we identify particles with
galaxies, both $\alpha$ and $\varphi$ contain information about
the similarity of the galaxy internal dynamics to the galaxy system
dynamics. In fact,
$$ \varphi \propto {\sigma_i\over\sigma} {p_{\rm min}\over r_c}
\eqno(4.1b)$$
and
$$\alpha \propto \left(\sigma_i\over\sigma\right)^4 {r_c\over R_s}
 \eqno(4.1c)$$
where $\sigma_i$ is the velocity dispersion of the stars
within an individual galaxy,
$r_c$ and $R_s$ are the galaxy and the system size, respectively, and
$p_{\rm min}$ is the maximum separation
at which two interacting galaxies are considered a merger remnant.

In order to test the hypotheses leading to equations (4.1),
we  follow the evolution of a King sphere (King 1966) with
Hernquist's (1987) $N$-body code. King spheres have a truncated
Gaussian velocity distribution and are stable self-gravitating systems.
Therefore, they approximate the boundary conditions of our physical model.

We sample two King spheres with
central gravitational potential $\phi(0)/\sigma^2=-12$.
The system sphere contains $N_g=50$ particles; the galaxy
sphere contains $N_s=100$ particles. We evolve each sphere in isolation
for $4.8$ collapse times $t_c$. We set the tolerance
parameter $\theta=0.8$ and the time step $\Delta t=10^{-3}t_c$. A softening
parameter $\varepsilon=0.1r_t$, where $r_t$ is the tidal radius of the
sphere, insures suppression of two-body relaxation effects.
As expected, the spheres are dynamically stable
and their velocity distributions remain remarkably Gaussian for the
entire integration.

We then replace each particle of the final system sphere with a 100 particle
final galaxy sphere. In other words, the 50 single particles become
resolved ``galaxies'' containing 100 particles each.
We  opportunely rescale particle velocities and
relative positions to ensure dynamical equilibrium
and  to suppress two-body relaxation within each galaxy.
This procedure is equivalent to switching on the particle internal
degrees of freedom. We evolve the system for $2.4t_c$. Simulation units
are $G=M=R=1$, where $G$ is the gravitational constant, $M$ the total
mass and $R$ the radius of the system. At each time, we identify
galaxies from particle positions through a generalization of the
friends-of-friends algorithm (Diaferio, Geller, \& Ramella 1994).
After $2.4t_c$ the galaxy number usually decreases from 50 to $\sim
30$.

We ran five simulations with different random number seeds.
Fig. 5 shows the time evolution of the distribution $\lambda(u)$ of
the galaxy pairwise velocity difference moduli.
Differences among the distributions of
the five simulations arise from statistical effects only.
Thus, in Fig. 5 we suppress statistical noise by summing the
distributions of the five simulations at each time.
Two fits are superimposed: the bold curve is the
perturbed Maxwellian distribution given in the Appendix (eqs. [A.12])
$$ \lambda(u)du = C(\sigma,\alpha,\varphi) u^2
\exp\left(-{u^2\over
4\sigma^2}\right) \left[1-\alpha H\left({u\over\sigma \sqrt{2}},
\varphi\right)\right]du \eqno(4.2) $$
and  the solid curve is the Maxwellian distribution, i.e. equation (4.2)
when $\alpha=0$.

\topinsert
\baselineskip=10pt
\moveleft 0.1\hsize
\vbox to .8\vsize {\hfil
\psfig{figure=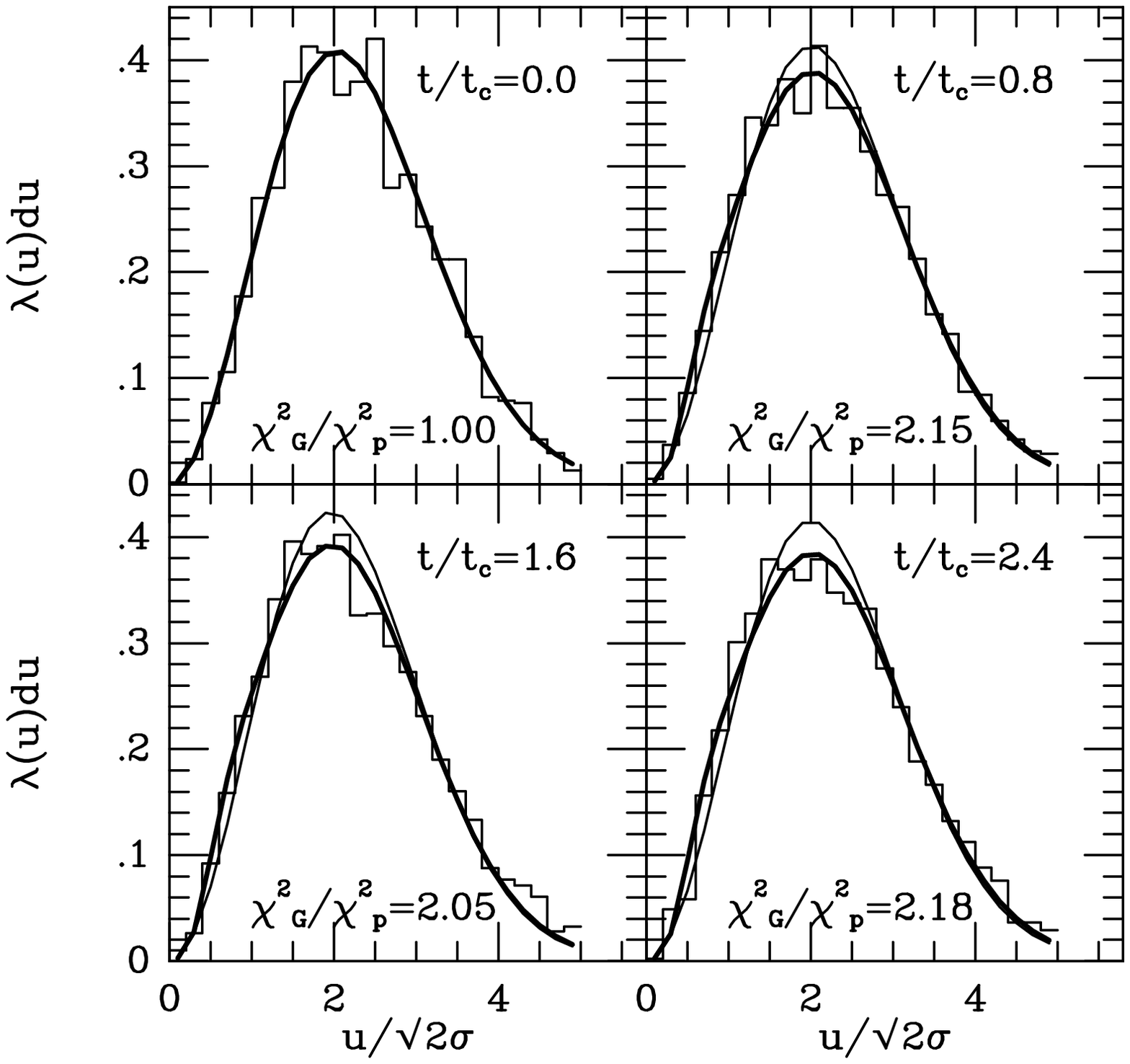,height=.6\vsize,width=0.8\hsize}\hfil
}
\vskip -0.4\vsize
\noindent
{\typenine {\bf Figure 5}
Time evolution of the distribution
of pairwise velocity difference moduli for systems of
particles with internal degrees of freedom. The distribution is
 initially Maxwellian but at later times the frequency of small relative
velocities increases. The best Maxwellian fits (solid lines) and the
equation
(4.2) fits (bold lines) are shown. The latter are slightly better than
the former as shown by the ratios of the $\chi^2$'s. In simulation
units, the Maxwellians
have fit parameters $\sqrt{2}\sigma = 1.43$, 1.42, 1.39, and 1.41 at
$t/t_c=0.0$, 0.8, 1.6, and 2.4, respectively. The perturbed Maxwellians
have fit parameters $\sqrt{2}\sigma = 1.44$, 1.45, 1.41, and 1.44 and
$\alpha = 0.01$, 0.13, 0.14, and 0.15, respectively.}
\endinsert

In the perturbed Maxwellian we set $\varphi=1.22$ according to equation
(A.4); $\sigma$ and $\alpha$ are free parameters. By adding
$\varphi$ as a free parameter the fits do not change significantly and
$\varphi$ remains in the range $1.00\div 1.40$.
Therefore, the perturbed Maxwellian only has
$\alpha$ as an additional free parameter compared with
the Maxwellian distribution. Fig. 5 shows that the perturbed
Maxwellian describes
the increased frequency of small relative velocities. However,
the frequency increase is barely detectable, despite the fact that
typically $\sigma_i/\sigma\sim 0.9$ (eq. [4.1c]); i.e. the ratio of the
velocity dispersions is not negligible. The
ratio of the $\chi^2$'s of the two distributions shows that the perturbed
Maxwellian fits the numerical distribution only slightly better
than the Maxwellian distribution.

At earlier times ($t\simless 2t_c$) most galaxies have not yet merged
and their masses have not been reduced significantly by tidal stripping.
Therefore, the physical model outlined in the Appendix is approximately
valid.
When we follow the system evolution for $t>2.4t_c$ we find that $\lambda(u)$
does not usually tend to depart farther from the unperturbed
distribution. At these later times however, comparison of equation (4.2)
with the numerical distributions is meaningless. Mergers create one
large merger remnant surrounded by galaxies with masses $\simless 0.1$
times the dominant galaxy mass. Thus, the system no longer contains
galaxies similar in mass and our simple physical assumptions break
down.

Thus, Fig. 5 confirms that $\Lambda(u,\sigma)$
in equation (4.1a) should be valid
for galaxy systems containing galaxies of similar mass. However, if
$\Lambda(u,\sigma)$ is Gaussian in the absence of galaxy internal degrees of
freedom we expect that small departures from a Gaussian distribution
will arise from the transfer of energy to the internal degrees of freedom.

In order to test equation (4.1a) against $\Lambda(u)$'s of real systems,
we compare equation (4.1a) with Hickson's compact groups (Hickson 1993).
These systems are the
densest in the Universe ($\sim 10^5$ galaxies per $h^3$ Mpc$^{-3}$)
if they are not two dimensional projections of
unrelated galaxies (Mamon 1992a; Hernquist, Katz, \& Weinberg 1995).
Therefore, we
expect that the kinetic energy loss effects might be detectable in these
extreme systems. Moreover, Hickson's brightness selection criterion
requires that galaxy members lie within an interval of three magnitudes,
assuring that galaxy members are not very different in mass.

\topinsert
\baselineskip=10pt
\moveleft 0.1\hsize
\vbox to .9\vsize {\hfil
\psfig{figure=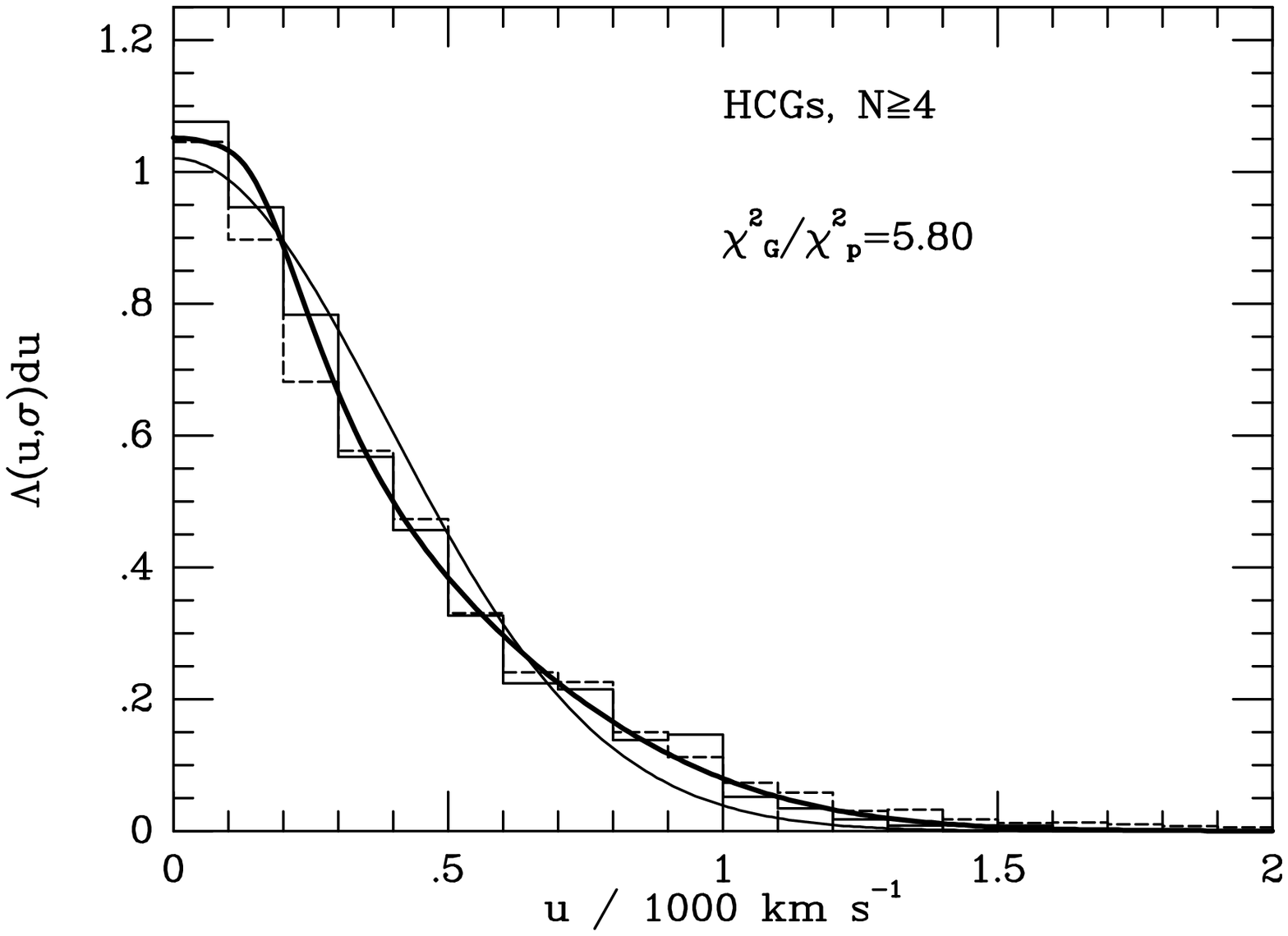,height=.6\vsize,width=0.8\hsize}\hfil
}
\vskip -0.53\vsize
\noindent
{\typenine {\bf Figure 6} Galaxy  pairwise velocity difference
distribution
for the 69 Hickson compact groups with $N\ge 4$ members. The solid line
is the best fit Gaussian and the
bold line is the best fit perturbed Gaussian (eq. [4.1a]). The
Gaussian has fit parameter $\sqrt{2}\sigma = 391$ km s$^{-1}$. Equation (4.1a)
has
fit parameters $\sqrt{2}\sigma = 487$ km s$^{-1}$ and $\alpha=0.69$. The dashed
histogram is the pairwise velocity difference distribution for the
compact groups with $N\ge 4$ members in the simulated catalog of
Diaferio {\etal} (1995).}
\endinsert

Fig. 6 shows $\Lambda(u)$ for the 69 Hickson (1993) compact groups with
$N\ge 4$ galaxies. We sum all
the single distributions for the whole sample of
69 compact groups because we assume that each compact group is a sample of the
same parent distribution. We base this approach on the model first
proposed by  Diaferio, Geller, \& Ramella (1994, 1995)
that observed compact groups
may be identified with substructures in collapsing rich loose groups.
Ramella {\etal} (1994) search the redshift neighborhoods of compact
groups within the CfA magnitude limited redshift surveys
 and confirm that at least 70\% of compact groups are
embedded in larger systems. If compact groups share this same origin, we
may assume that the pairwise velocity distribution
of each compact group is sampled from the same parent distribution.

Equation (4.1a) (bold curve) clearly fits the observed distribution better
than a Gaussian distribution (solid curve).
This result indicates that compact group galaxies are loosing kinetic energy
in a way consistent with our simple physical model. Moreover,
$\Lambda(u,\sigma)$
is only slightly perturbed compared with a Gaussian, indicating that
the galaxies still retain most of their orbital kinetic energy.
This conclusion agrees with the hypothesis that compact groups have just
collapsed and that most galaxies are at their first encounter within the
compact group (Diaferio {\etal} 1994).

Fig. 6 shows a further confirmation of the validity of this model of the
formation of compact groups. The dashed histogram is
the pairwise velocity distribution computed from the simulated
catalog of compact groups with $N\ge 4$ members (Diaferio {\etal} 1995).
The Kolomogorov-Smirnov test shows
that the observed compact group sample and the simulated
sample belong to the same parent distribution at the 16\% significance
level.

However, the difference between the two distributions is not a
statistical effect: the $N$-body
model systematically underestimates the frequency of
small relative velocities. This
result suggests that the model dynamical resolution is insufficient to
resolve the transfer of kinetic energy completely. In other words, the
model tends to merge galaxies before Nature does. This result is expected:
the model only accounts for dissipationless galaxy formation processes.
Dissipative processes decrease the galaxy merging cross-sections
and galaxies survive for a longer time against merging than in dissipationless
$N$-body simulations (e.g. Evrard, Summers, \& Davis 1994; Frenk {\etal} 1996).

\topinsert
\baselineskip=10pt
\moveleft 0.1\hsize
\vbox to .9\vsize {\hfil
\psfig{figure=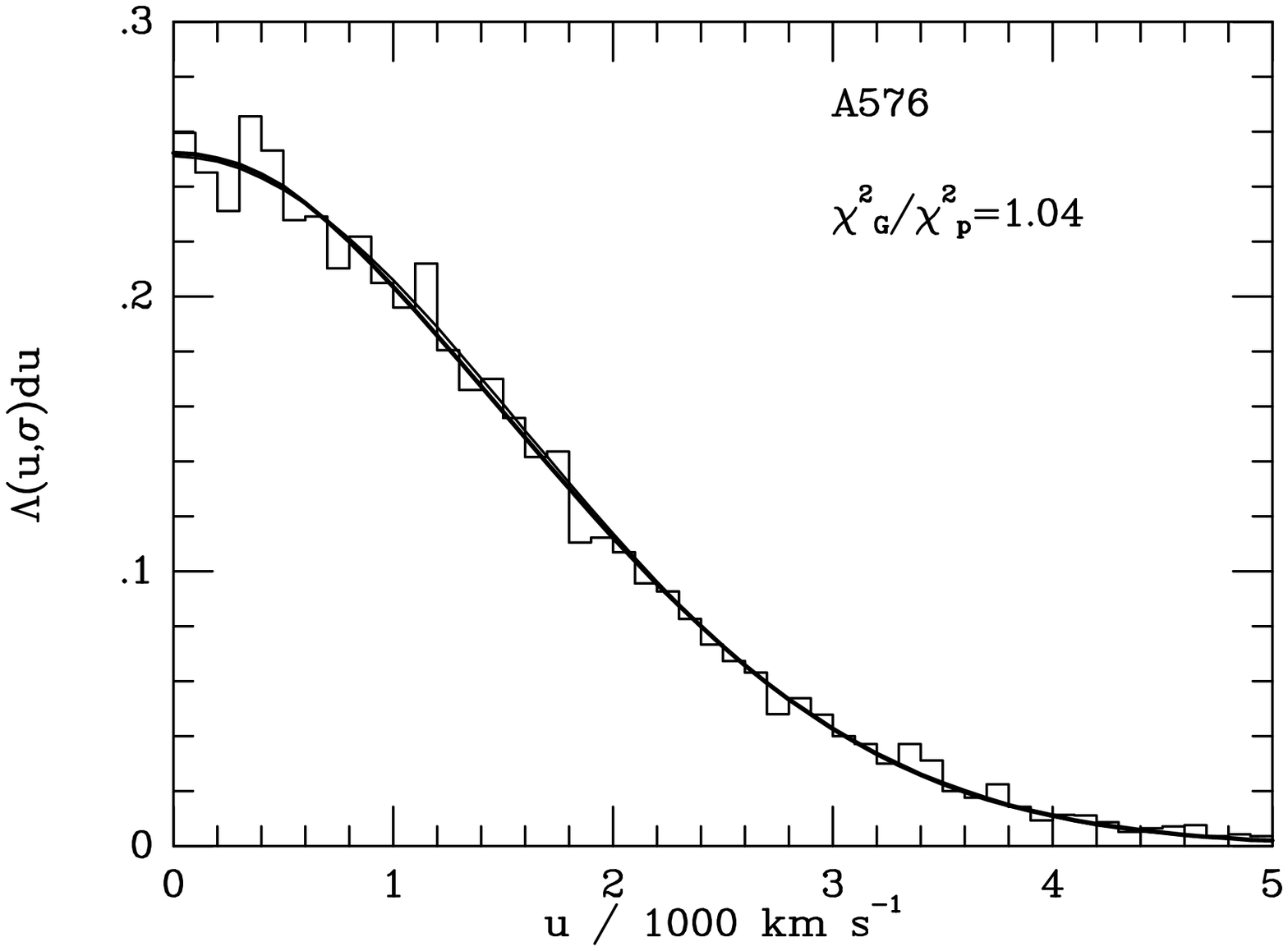,height=.6\vsize,width=0.8\hsize}\hfil
}
\vskip -0.53\vsize
\noindent
{\typenine {\bf Figure 7} Galaxy  pairwise velocity difference
distribution for the central region of the
Abell cluster A576. The solid line is the best fit Gaussian and the
bold line is the best fit perturbed Gaussian (eq. [4.1a]). The ratio of
the $\chi^2$'s shows that the two fits do not differ. The
Gaussian has fit parameter $\sqrt{2}\sigma = 1589$ km s$^{-1}$. Equation (4.1a)
has
fit parameters $\sqrt{2}\sigma = 1607$ km s$^{-1}$ and $\alpha=0.05$.}
\endinsert

We finally consider the pairwise velocity distribution for a galaxy cluster.
The perturbed distribution in equation
(4.1a) depends on the fourth power of the
ratio between the internal velocity dispersion of individual galaxies
and the velocity dispersion of the galaxies within the cluster (eq.
[4.1c]). Thus, we expect a nearly Gaussian $\Lambda(u,\sigma)$ for a massive
virialized cluster. We consider the Abell cluster A576 (Mohr {\etal} 1996).
The 85\% complete magnitude limited sample contains 169 galaxies lying
within a projected distance $r\simless 1.5 h^{-1}$ Mpc.
The cluster mass lies in the range $\sim 1\div 4 \times 10^{15} h^{-1} M_\odot$
implying a turnaround radius $\sim 3.0h^{-1}$ Mpc. Therefore,
the infall region around the turnaround radius
is not sampled. Of the 169 galaxies within the central region,
58 galaxies have
spectra with line emission and 111 have no line emission. Mohr {\etal}
(1996) identify these two samples with galaxies containing or not containing
star formation regions, respectively. They also generically identify them
with late-type or early-type galaxies.  Mohr {\etal} (1996) show
that the late-type galaxies have a velocity distribution broader than the
early-type galaxies and
identify late-type galaxies with galaxies falling into
the central region for the
first time. Thus, if we exclude these galaxies, the subsystem of
early-type galaxies is in approximate virial equilibrium.
Mohr {\etal} (1996) argue that the absence of apparent substructures in
the distribution of the early-type galaxies confirms this assumption.
Fig. 7 shows the pairwise velocity distribution for the
non-emission line galaxies only. In contrast with the compact group
sample where $\sigma_i/\sigma\sim 1$, here we have $\sigma_i/\sigma\sim
0.3$ (eq. [4.1c]). We fit the distribution with $\sigma$
and $\alpha$ free parameters; $\varphi=1.22$ as for the numerical
experiment.  As expected from equation (4.1c), the Gaussian and equation
 (4.1a) fit the observed distribution equally well.

\blankline
{\bf 5. THE REDSHIFT-SPACE CORRELATION FUNCTION}

\nobreak
In the preceding sections we investigate how the
exponential shape of the PVDF depends on the galaxy system number
density $n(\sigma)$ and the internal galaxy pairwise velocity distribution
$\Lambda(u,\sigma)$ of individual systems. We assume that all the galaxy
systems have the same $\Lambda(u,\sigma)$.
We then investigate how $\Lambda(u,\sigma)$ varies depending on the internal
dynamics of each system.

We now investigate how the redshift space correlation function $\xi_s$ depends
on the galaxy system number density $n(\sigma)$ and $\Lambda(u,\sigma)$
through the PVDF.
We restrict ourselves to Gaussian and exponential $\Lambda(u,\sigma)$'s.
We do not discuss the $\Lambda(u)$ distribution
for dissipative systems (Sect. 5), because this distribution is
only marginally distinguishable from a Gaussian for real systems.

The ``convolution method'' (e.g. Fisher 1995)
expresses the redshift-space correlation function $\xi_s$ as
$$ 1 + \xi_s(r_p,\pi) = \int_{-\infty}^{+\infty} [1+\xi(r)]p(u) dy
\eqno(5.1)$$
where $r_p$ is the spatial separation of the galaxy pair projected on
the sky, $\pi$ is the velocity difference along the line of sight,
$r^2 = r_p^2 + y^2$, $y$ is the pair spatial separation
parallel to the line of sight, and $u = \pi-y$ is the peculiar relative
velocity;\footnote{$^2$}{Here we assume that the velocities are
in units of the present Hubble constant $H_0$.}
$\xi(r)$ is the real-space correlation function and $p(u)$ is the PVDF.
In equation (5.1) we assume (1) that the PVDF is independent of $r$
and (2) that the mean relative peculiar velocity $v_{12}(r)$
of galaxy pairs separated by
$r$ is zero. Both assumptions are reasonable when $0.1h^{-1}$ Mpc
$\simless r_p\simless 1h^{-1}$
Mpc, where galaxy velocities are almost random ($v_{12}(r)\sim 0$) and
the pairwise velocity dispersion $\sigma_{12}(r)\sim {\rm const}$
(Marzke {\etal} 1995; Fisher 1995).

\topinsert
\baselineskip=10pt
\moveleft 0.1\hsize
\vbox to .8\vsize {\hfil
\psfig{figure=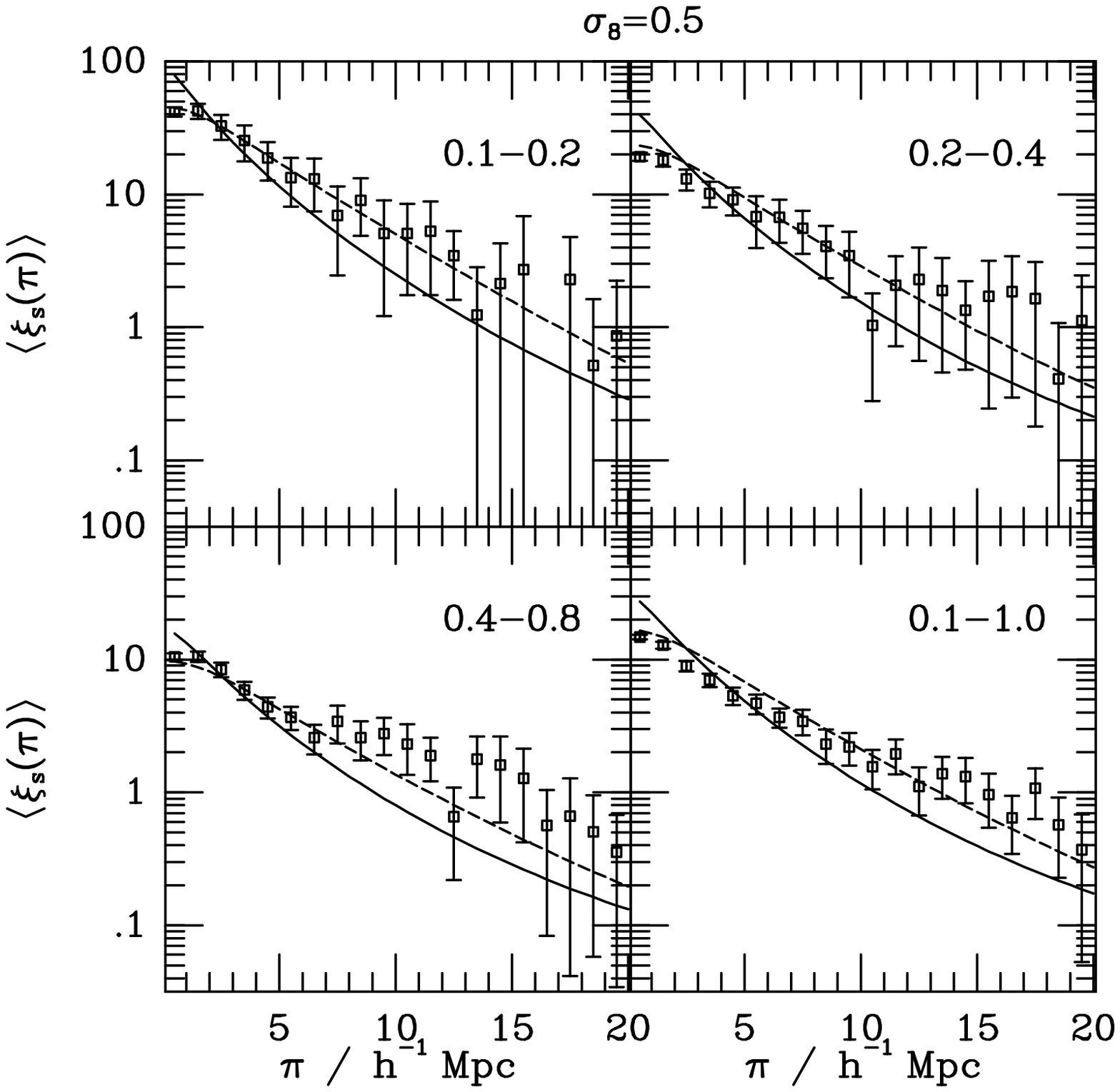,height=.6\vsize,width=0.8\hsize}\hfil
}
\vskip -0.4\vsize
\noindent
{\typenine {\bf Figure 8a} Comparison of the observed redshift-space
correlation function $\langle\xi_s(\pi)\rangle$
with models of the PVDF convolved with the observed
real space correlation function (eqs. [5.2] and [5.1]).
Squares are the measures from the CfA2+SSRS2 redshift survey
samples with $[r_{\rm min},r_{\rm max}]=[0.1,0.2]$, $[0.2,0.4]$,
$[0.4,0.8]$, and $[0.1,1.0] h^{-1}$ Mpc respectively, as shown in each
panel
(Marzke {\etal} 1995). Curves
are the PVDFs (eq. [2.1]) with $\sigma_{\rm min}=100$ km s$^{-1}$ and
$\sigma_{\rm max}=1500$ km s$^{-1}$. Solid (dashed) lines are
computed with an exponential (Gaussian) internal pairwise
velocity distribution $\Lambda(u,\sigma)$. The number density of galaxy
system is the observed $n(\sigma)$ (eq. [2.4]) or the
Press-Schechter $n(\sigma)$ for a flat CDM universe (eq. [2.3a])
with different normalization $\sigma_8$: (a) $\sigma_8=0.5$; (b)
$\sigma_8=1.0$; (c) $\sigma_8=1.5$; (d) observed $n(\sigma)$.}

\baselineskip=10pt
\moveleft 0.1\hsize
\vbox to .8\vsize {\hfil
\psfig{figure=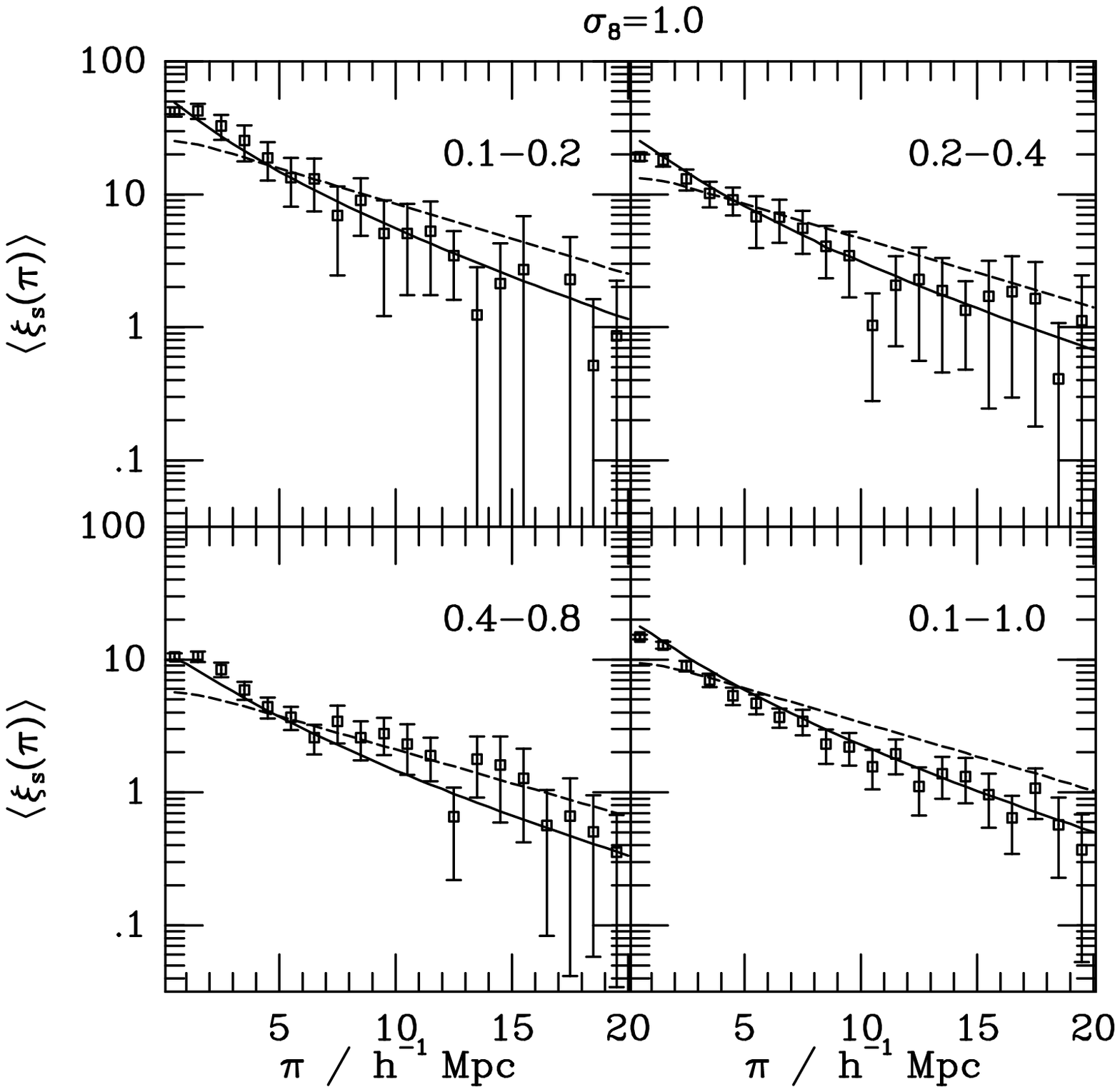,height=.6\vsize,width=0.8\hsize}\hfil
}
\vskip -0.4\vsize
\noindent
{\typenine \centerline{\bf Figure 8b} }
\endinsert

\topinsert
\baselineskip=10pt
\moveleft 0.1\hsize
\vbox to .8\vsize {\hfil
\psfig{figure=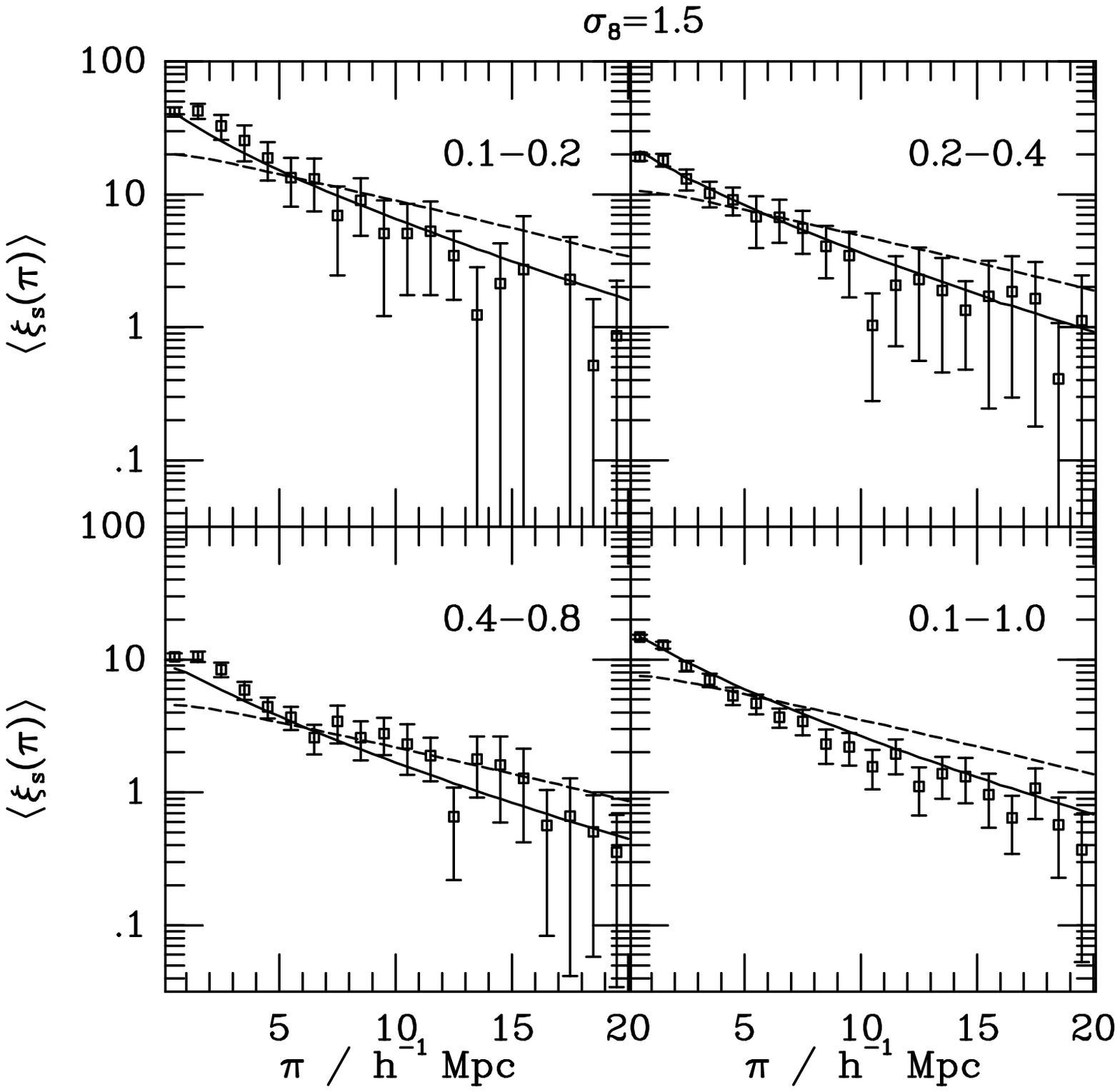,height=.6\vsize,width=0.8\hsize}\hfil
}
\vskip -0.4\vsize
\noindent
{\typenine \centerline{\bf Figure 8c} }

\baselineskip=10pt
\moveleft 0.1\hsize
\vbox to .8\vsize {\hfil
\psfig{figure=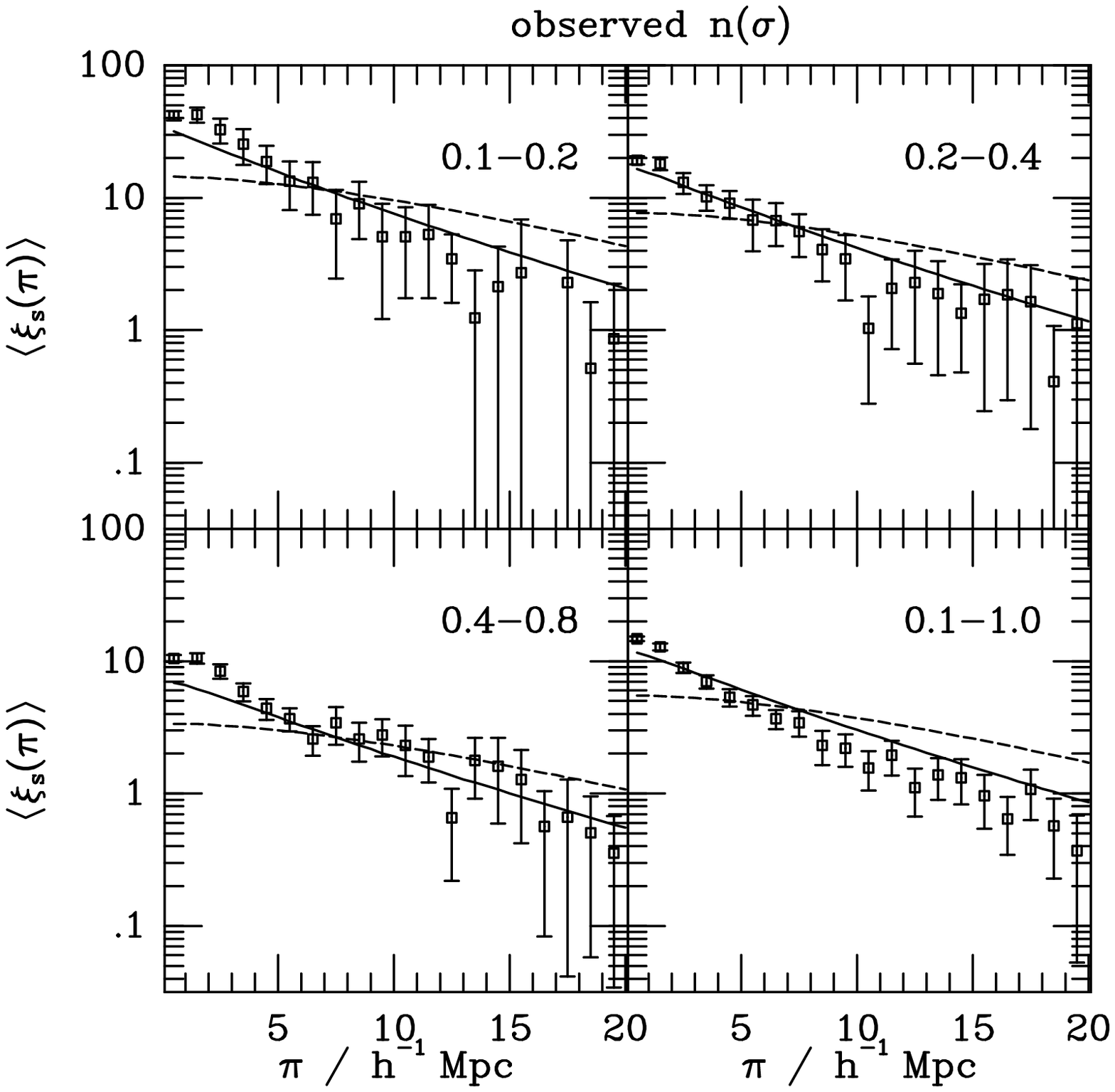,height=.6\vsize,width=0.8\hsize}\hfil
}
\vskip -0.4\vsize
\noindent
{\typenine \centerline{\bf Figure 8d} }
\endinsert

We consider the redshift-space
correlation function averaged over the projected separation $r_p$, namely
$$ \langle  \xi_s(r_{\rm min},r_{\rm max},\pi)\rangle
= {1\over r_{\rm max}-r_{\rm min}}
\int_{r_{\rm min}}^{r_{\rm max}} \xi_s(r_p,\pi) dr_p. \eqno (5.2)$$

Fig. 8 shows the comparison of the measured
$\langle \xi_s(\pi)\rangle$ for different intervals of projected
separations $[r_{\rm min},r_{\rm max}]$
with different models of the PVDF convolved
with the real space correlation function $\xi(r)$. We use $ \xi(r) =
(r/r_0)^\gamma$, where $r_0=5.97\pm0.15 h^{-1}$ Mpc and $\gamma=-1.81\pm
0.02$ as measured by Marzke {\etal} (1995) for the CfA redshift survey
(CfA2) and the Southern Sky Redshift Survey (SSRS2)
galaxy samples combined (CfA2 + SSRS2). In Fig. 8, squares
are the measured $\langle \xi_s(\pi)\rangle$ for this sample with
$[r_{\rm min},r_{\rm max}]=[0.1,0.2]$, $[0.2,0.4]$, $[0.4,0.8]$, and
$[0.1,1.0]h^{-1}$ Mpc.
We superimpose the curves computed through the integrals
in equations (5.2) and (5.1), where $p(u)$ is the integral in equation
(2.1) with $\sigma_{\rm min}=100\kms$ and $\sigma_{\rm max}=1500\kms$.
We show the curves with
a Gaussian internal velocity distribution $\Lambda(u,\sigma)$ (dashed
lines) and with an exponential $\Lambda(u,\sigma)$ (solid lines). We compute
the
curves with both the observed $n(\sigma)$ (eq. [2.4]) and with the distribution
derived from the Press-Schechter theory (eq. [2.3a]) for $\sigma_8=0.5$,
1.0, and 1.5.

In order to quantify the agreement between the curves and the data for
each projected separation interval, we
plot the $\chi^2$ per degrees of freedom $\nu$ as a function of
the upper limit $r_{\rm max}$ of the interval
(Fig. 9).  We do not derive a parameter from the twenty data points
within each range of $r_p$, thus we assume $\nu=20$.
The $\chi^2$'s are indicative and are meaningful only if we compare them
with each other. The data are actually correlated and the estimates of
$\langle \xi_s(\pi)\rangle$ are not normally distributed. Therefore, we should
use a different approach to estimate $\chi^2$ (see Fisher {\etal}
1994a; and Marzke {\etal} 1995 for further details).

The observed $n(\sigma)$ approximates the data better when it weights an
exponential $\Lambda(u)$ (Fig. 9, open squares) rather than a Gaussian
(filled squares). However, systems selected from a redshift survey exceed a
particular density contrast threshold $\delta\rho/\rho$. The $n(\sigma)$ we use
thus contains only systems with $\delta\rho/\rho\ge 80$ with respect to the
background (Zabludoff {\etal} 1993b). Therefore,
the observed $n(\sigma)$ does not contain enough systems with small density
contrast, and presumably low $\sigma$, by definition.
The exponential $\Lambda(u)$ partially compensates this
underestimate and the agreement is better.

In any case, the theoretical $n(\sigma)$ confirms that an exponential
$\Lambda(u)$ reproduces the data better than the
Gaussian $\Lambda(u)$, although not for all the
normalizations $\sigma_8$. The theoretical $n(\sigma)$ also depends on
the power spectrum $P(k)$ (eq. [2.3c]) and on
the density of the Universe through the perturbation growth factor
which enters the Press-Schechter distribution function.
Thus, we must interpret the implications of Fig. 9 about
$\sigma_8$ cautiously.

\topinsert
\baselineskip=10pt
\moveleft 0.1\hsize
\vbox to .8\vsize {\hfil
\psfig{figure=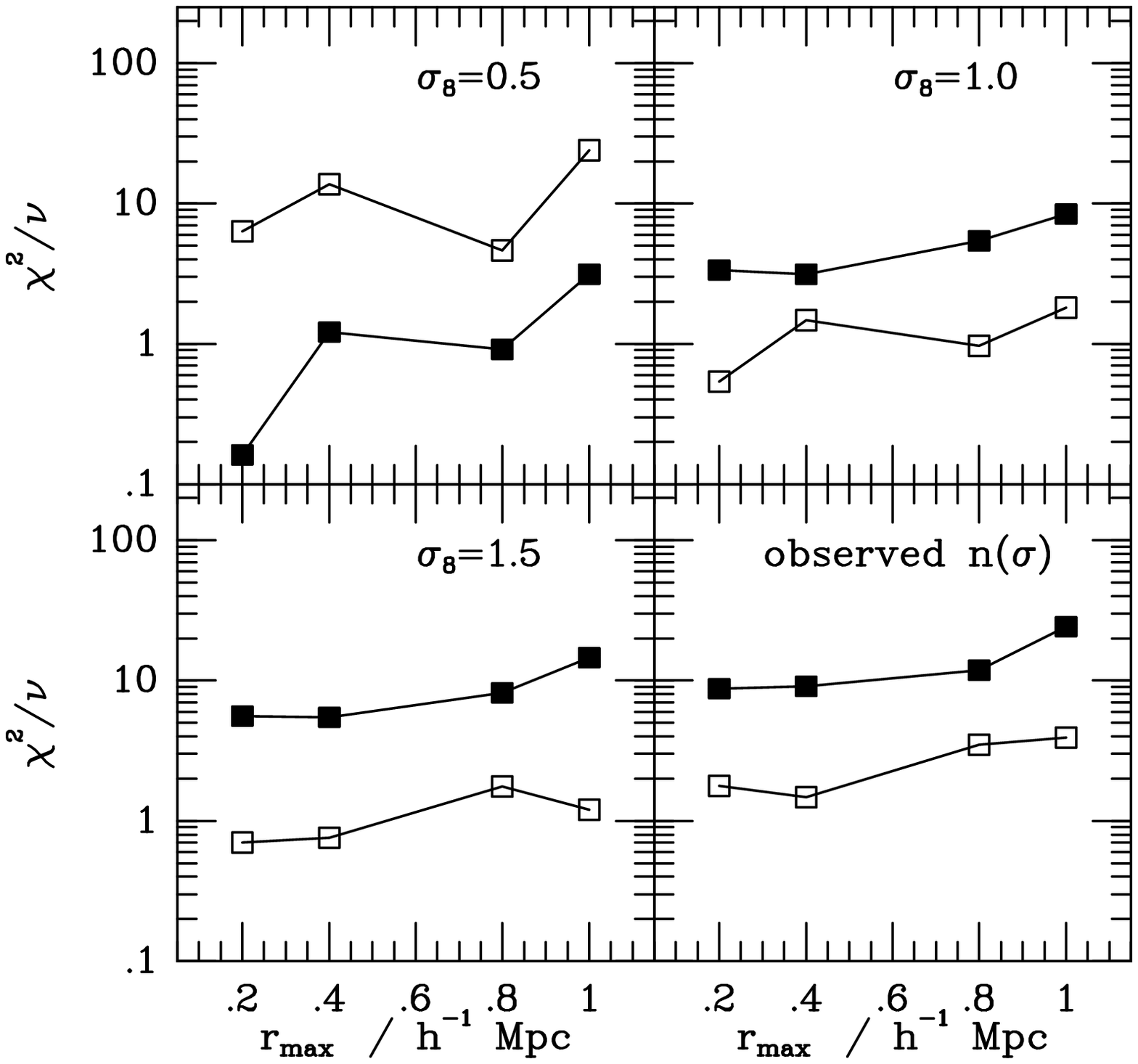,height=.6\vsize,width=0.8\hsize}\hfil
}
\vskip -0.4\vsize
\noindent
{\typenine {\bf Figure 9} Reduced $\chi^2$ with $\nu=20$ degrees of
freedom for the curves in Fig. 8  as a function of the upper limit
$r_{\rm max}$
of the integral in equation (5.2). The four upper limits correspond to
the four projected separation intervals $[r_{\rm min},r_{\rm
max}]=[0.1,0.2]$, $[0.2,0.4]$, $[0.4,0.8]$, and $[0.1,1.0] h^{-1}$ Mpc.
Open (filled) squares are for an exponential (Gaussian)
$\Lambda(u,\sigma)$.}
\endinsert

The agreement between curves and observations also depends
on the projected separation interval.
Marzke {\etal} (1995) computed $\langle \xi_s(\pi)\rangle$ for the data with
the intervals shown in Fig. 8. However, the
galaxy coordinates in the Zwicky catalog, on which the CfA survey is
based, are accurate to $\sim 1\div 1.5$ arcmin. At redshift
$cz=10000\kms$, the $3\sigma$ error is thus $\sim 0.09\div 0.14h^{-1}$
Mpc, implying an error in the projected separation $\sim 0.13\div
0.19h^{-1}$ Mpc. With the current data, large errors probably contaminate
the interval $[r_{\rm min},r_{\rm max}] = [0.1,0.2] h^{-1}$ Mpc.
Intervals with $r_{\rm min}\ge 0.2h^{-1}$ Mpc are more reliable.  We also
emphasize that galaxy velocities often have uncertainties $\simgreat 50\kms$,
which means typical errors $\simgreat 70\kms$ in the relative
velocities. Thus, we regard the measures of $\xi_s$
at $\pi=50$, $150\kms$ with caution.

Large differences in $\chi^2$'s
between the exponential and the Gaussian $\Lambda(u,\sigma)$ (see for
example Fig. 9 when $\sigma_8=0.5$ or when $\sigma_8=1.0$ and $[r_{\rm
min},r_{\rm max}]=[0.1,0.2]$ or $[0.4,0.8]h^{-1}$ Mpc) originate
mainly at small relative velocities $\pi\simless 200\div 300\kms$ (Fig.
8). Thus, by improving  the accuracy of galaxy relative positions and
velocities by at least a factor of two, the redshift-space correlation
function $\xi_s$ can (1) clearly discriminate
among the models, and (2) can probably separate the cosmological
contribution represented by $n(\sigma)$ from the contribution of
the internal dynamics of galaxy systems represented by
$\Lambda(u,\sigma)$.
Moreover, on very small scales ($r_p\simless 0.1\div 0.2h^{-1}$ Mpc,
$\pi\simless 100\div 200\kms$), we expect that the dissipative effects
outlined in Sect. 4 will become more apparent. Thus, $\xi_s$
can constrain the importance of mergers in the present Universe.

The main conclusion of our analysis is that
agreement between the models of the exponential PVDF and the observed
redshift-space correlation function on non-linear scales
depends strongly on both the underlying cosmogonic model [namely
$n(\sigma)$] and the internal dynamics of galaxy systems [i.e.
$\Lambda(u,\sigma)$]. Neither aspect dominates.
However, reliable measures of $\xi_s$ at small scales can separate the
two contributions and provide further constraints on the model of the
Universe.

\blankline
{\bf 6. CONCLUSION}

\nobreak
Marzke {\etal} (1995) measured the redshift-space correlation
function $\xi_s$ for galaxy samples of the Center for Astrophysics (CfA)
redshift surveys for galaxy separations $\simless 1h^{-1}$ Mpc.
An exponential galaxy pairwise velocity distribution
function (PVDF) yields the best fit. This result is common to other
redshift surveys (e.g. Bean {\etal} 1983; Fisher {\etal} 1994b).

We propose a physical explanation for this observed exponential shape.
If all galaxies belong to isolated galaxy systems
with velocity dispersion $\sigma$,
the PVDF is the weighted sum of the distributions $\Lambda(u,\sigma)$
of the pairwise velocities $u$ within each system.
The weight depends on the galaxy number $\nu(\sigma)$ within
each system and the number density $n(\sigma)$ of the systems within the
sample.

We assume that $\Lambda(u,\sigma)$ is a universal function, identical for
each system. This assumption is inadequate, because we show that the
shape $\Lambda(u,\sigma)$ depends on the dynamical state of the system.
However, if we assume that all the system are virialized, $\Lambda(u)$
is Gaussian and $\nu(\sigma)\propto \sigma^2$. In this case,
both the observed $n(\sigma)$
and the $n(\sigma)$ predicted by the Press-Schechter theory in a flat
CDM Universe yield a nearly exponential PVDF,
but only at large relative velocities $u$.
In order to obtain an exponential central peak, $\Lambda(u)$
has to be more centrally peaked than a Gaussian distribution. When a galaxy
system is unrelaxed, substructures and infall regions contribute
to a centrally peaked $\Lambda(u)$. We limit our analysis to an exponential
$\Lambda(u,\sigma)$ which yields the expected central peak of the PVDF.
The Gaussian and the exponential distributions
represent the two limiting cases. A more detailed analysis of the
physical origin of $\Lambda(u,\sigma)$ is likely to yield a
$\Lambda(u,\sigma)$ between these two cases. Therefore,
we conclude that the observed
exponential PVDF testifies to the presence of a large fraction of
unrelaxed galaxy systems in the present-day Universe.

A third process may increase the frequency of small relative velocities:
the transfer of orbital kinetic energy to galaxy internal degrees of
freedom. We derive an analytical $\Lambda(u)$ which accounts for
energy transfers driven by tidal perturbations. We predict that
these perturbations are
detectable in galaxy systems with a ratio $\simgreat 1$ between the internal
velocity dispersion of individual galaxies and the velocity dispersion of
galaxies within the system. We confirm this prediction by comparing the
analytic distribution with $N$-body simulations and with observed
compact groups.

Finally, we compare the measured redshift-space correlation function
$\xi_s$ with the convolution of different models of the exponential
PVDF with the measured real-space correlation function.
The agreement between models and observations
depends strongly on both the underlying cosmogonic model and the
internal velocity distribution $\Lambda(u,\sigma)$ of galaxy systems.
These two effects are of comparable importance
over the entire range of relative velocities $\pi<2000\kms$.

We expect to be able to disentangle
the two effects with more accurate galaxy coordinate and relative
velocity measurements. In fact, the redshift-space
correlation function $\xi_s$ at projected separations $r_p\simless
0.5 h^{-1}$ Mpc and relative velocities $\pi\simless 300\kms$
is very sensitive to the shape of $\Lambda(u,\sigma)$.
Thus, a better measure of $\xi_s$ at very small scales
 poses strong constraints on the shape of
$\Lambda(u,\sigma)$ and will improve our understanding of the dynamics of
galaxy systems on these scales.

After the submission of this paper, we learned of Sheth's (1996)
independent work on the problem of the exponential shape of the PVDF.
He investigates a model similar to the model we outline in Sect. 2.
He performs an accurate comparison of this
model with $N$-body simulations and shows that for initial density
perturbations with  power-law power spectra the PVDF is well
approximated by an exponential.  He shows that the assumptions which
underlie our analytic approach in Sect. 2 include the relevant physics.

\blankline
We sincerely thank Ron Marzke for suggesting the problem of the
pairwise velocity distribution and for intensive electronic
correspondence.
We are grateful to Joe Mohr for providing us with galaxy redshifts of the
Abell cluster A576 in advance of publication.
We thank Ira Wasserman and Simon White for pointing out
some aspects of the internal velocity distributions of galaxy
systems which were not adequately treated at an earlier stage, and
Bhuvnesh Jain for identifying some initially obscure statements. We
thank an anonymous referee for the clarifying suggestions of a prompt report.
Invaluable long discussions with Paola Ciarpallini and her inexhaustible
support made this work possible. We warmly dedicate this work to Paola.
This research is supported in part by NASA Grant No. NAGW-201 and by the
Smithsonian Institution. A.D. was a Center for Astrophysics Pre-Doctoral
Fellow.

\vfill\eject
\blankline
\centerline {\bf APPENDIX}

\blankline
\nobreak
Here we derive the pairwise velocity difference distributions
given in equations (4.1) and (4.2) for dissipative systems.

 We assume a self-gravitating gas
of particles with an initial Gaussian velocity distribution.
The particles have internal degrees of freedom; tidal effects
increase the particle internal
energy during particle motion within the system at the expense of
the orbital kinetic energy of the particles. In the following derivation
we ignore particle  mass loss
and therefore we underestimate the total kinetic energy loss as
discussed in Sect. 4.

First we compute the fraction of the relative kinetic
energy transferred into the particle internal degrees of freedom.
We then derive an approximated pairwise velocity difference
distribution.

Suppose that the particles have equal mass $m$. $E$ is
the relative kinetic energy per unit mass of two particles
in their center of mass reference frame. If $v$ is their relative
velocity we have $E=v^2/4=\sigma_s^2 u^2/2$ where $\sigma_s$ is
the one-dimensional velocity dispersion of the system
and $u=v/\sqrt{2}\sigma_s$.

Spitzer (1958) considered the encounter between
a cloud and a star cluster and computed the relative energy transfer to
the internal energy of the cluster during
the encounter. The derivation is similar to the
derivation of the energy transfer in a Coulomb collision between
a moving charge and a harmonically
bound charge in the dipole approximation (Jackson 1962). In fact,
Spitzer
assumed that for the stars in the cluster (1) the tidal force of the
cloud is small compared with the gravitational attraction of the cluster
and (2)
that the internal gravitational potential of the cluster is proportional
to the
square of the distance from the cluster center, i.e. the stars are
harmonic oscillators with the same frequency.

In Spitzer's paper the cloud and cluster have mass  $m_n$ and $m_c$
respectively, $r_c^2$ is the
mean square cluster radius, $1/\omega$ the oscillation period of the
stars in
the cluster, $v$ the cloud-cluster relative velocity and $p$ the impact
parameter.  Spitzer showed that the increase of the cluster
internal energy after a single encounter is
$$ \Delta E = {m_cr_c^2\over 6} \left(2Gm_n\over vp^2\right)^2
L\left(2\omega p\over v\right) \eqno (A.1a)$$
where $G$ is the gravitational constant and
$$ L(\theta) = 2\theta^2[\theta^2K_0^2(\theta)+ \theta
K_0(\theta)K_1(\theta) + (1+\theta^2)K_1^2(\theta)]\eqno(A.1b)$$
where $K_0$ and $K_1$ are the usual modified Bessel functions, and
$\theta=2\omega p/v$.

When $v\to \infty$, $L(\theta)\to 2(1+2\theta^2)\to 2$ and equation (A.1a)
reduces to the usual impulse approximation. When $v\to 0$,
$L(\theta)/v^2 \to \pi\theta^3\exp(-2\theta)/v^2\to 0$; the
energy loss falls exponentially to zero. Weinberg (1994a, b, c) shows
that such an adiabatic cutoff is not correct in general. In fact,
$\Delta E$
approaches zero when $\omega/v\to \infty$. However, a stellar system
always has stars with arbitrary small $\omega$ and for those stars we
never have $\omega/v\to \infty$. The perturbation suffered by those
stars ultimately affects the internal dynamics of the whole stellar
system. Thus, assuming $\Delta E\sim 0$ when $v\to 0$ will underestimate
the
energy loss. However, we would detect such energy loss only on time
scales
$\simgreat R/v$ where $R$ is some characteristic size of the system. In
other words, we underestimate the energy loss only if we observe
the system for a sufficiently long time interval.

We apply equations (A.1) to galaxy systems
where the adiabatic approximation does not break down because we do not
observe them for a long enough time. In fact,
either galaxy systems  are dynamically young
because they suffer a merging instability and are therefore intrinsically
unstable (groups) or their size is large enough that when $v$ is small,
$R/v$ is greater than the Hubble time (clusters). Therefore,
we expect that equations (A.1) approximate the energy loss for galaxies
within groups and clusters.

Let us now apply
equations (A.1) to a system of equal mass galaxies. All possible galaxy pair
combinations reduce to a ``reduced'' galaxy with mass $m_n=m/2$ (the
``cloud'')
which moves with velocity $v$ through a field of fixed sources with mass
$m_c=2m$ (the ``clusters'').
The ``cloud'' loses a fraction of its kinetic energy at a rate given by
$\Delta E$ times the number of
collisions per unit time, integrated over the impact parameter
$p$ and the mass of the ``clusters'' $m_c$.
In other words, if $n(m_c)$ is the number
density of ``clusters'' with mass between $m_c$ and $m_c+dm_c$, we have
$$\eqalignno{ {dE\over dt} & = -\int \Delta E\times v \times 2\pi pdp \
\times n(m_c) dm_c\cr
& = v\times {m_n^2r_c^2\over 6}\left(2G\over v\right)^2   2\pi
\left(2\omega\over v\right)^2 \int_{\theta_{\rm min}}^\infty
{L(\theta)\over \theta^3} d\theta
\int m_c n(m_c) dm_c. & (A.2a) \cr
}$$

We assume that all the ``clusters'' have the same mass $m_c=2m$.
Thus, the ``cluster'' mass spectrum per unit volume is
$n(m_c) = n_0 \delta(2m-m_c)$, where $\delta$ is the usual Dirac delta
function.

The quantity $\theta_{\rm min}$ depends on the validity of assumption
(1)
above, namely  that the ``cloud'' exerts a tidal force on a ``cluster''
star which is small compared to the ``cluster'' attraction.
We can write
$$ \theta_{\rm min}  = {2\omega p_{\rm min}\over\sqrt{2} \sigma_s u}
\equiv {\varphi\over u}. \eqno (A.3)$$
The oscillation period $1/\omega$ of the star within the ``cluster'' is
roughly twice the inverse of the crossing time $r_c/\sqrt{3}\sigma_i$
where
$\sigma_i$ is the one-dimensional velocity dispersion within the
``cluster''.
Therefore
$$ \varphi = {\sqrt{6}\over 2} \left(\sigma_i\over \sigma_s\right)
 {p_{\rm min}\over r_c}. \eqno (A.4)$$
For $p_{\rm min}\sim r_c$ and $\sigma_i\sim \sigma_s$ we expect $\varphi
= \sqrt{6}/2\sim 1.22$.

Applying the recursion formulas $dK_0/d\theta = -K_1$ and $dK_1/d\theta
= -K_0-K_1/\theta$ the indefinite integral over $\theta$ is
$$ -\int {L(\theta)\over \theta^3} d\theta = 2\theta
K_0(\theta)K_1(\theta) +K_1^2(\theta) \equiv  \tilde
L(\theta).\eqno(A.5)$$
Finally, equation (A.2a) reduces to
$$ {dE\over dt} = -{8\pi\over 3} m r_c^2 n_0 \omega^2 (Gm)^2{1\over v^3}
\tilde L\left(\varphi\over u\right). \eqno(A.2b)$$
For the virial theorem $Gm = 3\sigma_i^2r_c$. Moreover, $\omega =
\sqrt{3}\sigma_i/2r_c$ and $n_0=3N/4\pi R_s^3$ where $R_s$ is the size
of the system and $N$ is the total number
of galaxies within the system. We can write $N=GfM_{tot}/Gm =
f(\sigma_s/\sigma_i)^2(R_s/r_c)$ where $fM_{tot}$ is the fraction of the
total mass concentrated in galaxies.
 The crossing time of the system is $t_{cr} =
 \sqrt{3}\sigma_s/R_s$, thus we have
$$ \eqalignno {{dE\over dt}& = -{m\sigma_s^2\over t_{cr}}
{9\sqrt{6}\over 8}
N \left(\sigma_i\over\sigma_s\right)^6\left(r_c\over R_s\right)^2
{1\over u^3}
\tilde L\left(\varphi\over u\right)\cr
& = -{m\sigma_s^2\over t_{cr}} {9\sqrt{6}\over 8} f
\left(\sigma_i\over\sigma_s\right)^4\left(r_c\over R_s\right) {1\over
u^3}
\tilde L\left(\varphi\over u\right)\cr
& \equiv -{m\sigma_s^2\over t_{cr}} \beta {1\over u^3}
\tilde L\left(\varphi\over u\right) & (A.2c)\cr}$$
where we have introduced the constant
$$ \beta = {9\sqrt{6}\over 8} f
\left(\sigma_i\over\sigma_s\right)^4\left(r_c\over
R_s\right).\eqno(A.2d) $$
The constant $\beta$ contains information about the size and the
internal velocity dispersion of galaxies compared with those of the
whole
system. It is
apparent that the energy transfer into galaxy internal degrees of
freedom is mainly sensitive to the ratio of the velocity dispersions.
The constant $\beta$ specifies when we can ignore the galaxy internal
degrees of freedom. We see that within galaxy groups the energy transfer
must be more clearly  detectable than in galaxy clusters.

The energy transfer rate per unit mass in unit of $\sigma_s^2$
may finally be written
$$ {dE\over dt} = -{1 \over t_{cr}}
\beta {1\over u^3} \tilde L\left(\varphi\over u\right). \eqno(A.6)$$

Suppose now that we know the relative kinetic energy distribution
$q(E)dE$
at time $t$.
We wish to compute the energy distribution $\lambda(E)dE$ at time
$t+\eta t_{cr}$, when $\eta\to 0$, so that $u\sim {\rm const}$.

Using equation (A.6), the relative kinetic energy per unit mass
in unit $\sigma_s^2$ at the time $t+\eta t_{cr}$ is,
to first order in $\eta$,
$$ \eqalignno {E(t+\eta t_{cr}) &= E(t) + {dE\over dt} \eta t_{cr} \cr
 &=E(t)- \eta\beta\Delta(u,\varphi) & (A.7a)\cr}$$
 where
$$ \Delta(u,\varphi) = {1\over u^3} \tilde L\left(\varphi\over u\right).
\eqno(A.7b) $$

Now, the probability density $\lambda(E)dE$ is
$$ \lambda(E)dE= q[G^{-1}(E)] {dG^{-1}(E)\over dE} dE \eqno(A.8)$$
where the inverse function $G^{-1}$ is defined through equations (A.7)
$$ E = G(E_0) = E_0-\eta\beta\Delta(E_0) \eqno (A.7c)$$
and $E(t)=E_0$.

In the limit $\eta \to 0$, we have to first order in $\eta$
$$ G^{-1}(E) = E + \eta\beta\Delta(E) \eqno(A.9)$$
and we may rewrite equation (A.8)
$$ \lambda(E)dE = q(E) \left\{1+\eta\beta\Delta(E)\left[{d\log
[q(E)]\over dE}
+ {d\log[\Delta(E)]\over dE}\right]\right\}dE. \eqno (A.10)$$

Equation (A.10) is valid only when $\eta\to 0$, i.e. for times close to $t$
when the
distribution $q(E)$ is known. We should obtain the distributions
$\lambda(E,t)dE$ at different
times $t$ by solving the correct Boltzmann equation specific for our
problem. However, we wish to have an analytic distribution to compare
with real systems. Therefore, we go further and use equation (A.10) {\it
tout-court} assuming $\alpha =\eta\beta$ is a free fit parameter.
We justify this assumption with the following argument: if
$q(E)=\lambda_0(E)$ at time $t_0$ and $\lambda(E)=\lambda_1(E)$
at time $t_1$, we again have
equation (A.10) at time $t_2$ where $q(E)\to \lambda_1(E)$ and $\lambda(E)\to
\lambda_2(E)$. To first order in $\eta$, $\eta\beta\to 2\eta\beta$.
Thus,
if we apply this iterative procedure,
the form of equation (A.10) does not change, but now the coefficient $\alpha$
tells us how far the system is dynamically from
the initial distribution $q(E)$. Moreover, the coefficient $\alpha$
is a measure (1) of the time scale of the relative
velocity change because of kinetic energy loss ($\eta$) and (2) of the
similarity of the galaxy internal dynamics to the galaxy system
dynamics ($\beta$).

If we assume a Maxwellian $q(E)$
$$ q(E) dE \propto E^{1/2} e^{-E}  dE\eqno (A.11)$$
equation (A.10) becomes, in terms of the velocity modulus
$u=v/\sqrt{2}\sigma_s=(2E)^{1/2}$ and with the explicit expression of
$\Delta(u)$ (eq. [A.7b]),
$$ \lambda(u)du\propto u^2\exp\left(-{u^2\over 2}\right)
[1-\alpha H(u, \varphi)]du \eqno(A.12a)$$
where
$$ H(u, \varphi) = {1\over u^3}\tilde L\left(\varphi\over u\right)
  \left\{1+{2\over u^2} + {\varphi\over u^3} {d\log[\tilde
L(\theta)]\over d\theta}\right\}. \eqno(A.12b)$$

The one-component velocity density distribution $\Lambda(u)du$ is
related
to the density distribution of the velocity moduli through
the equation (Feller 1966)
$$ \lambda(u)du = - u {d\Lambda(u)\over du}. \eqno(A.13)$$
Equation (A.13) holds for any isotropic three-dimensional random field.
With the boundary condition $\Lambda(u)\to 0$ when $u\to \infty$ we
obtain
$$ \Lambda(u)du \propto \int_u^\infty {\lambda(t)\over t} dt.
\eqno(A.14)$$
In other words,
$$ \Lambda(u)du \propto \exp\left(-{u^2\over 2}\right)
[1-\alpha\tilde H(u,\varphi)] du\eqno (A.15a)$$
where
$$ \tilde H(u,\varphi) = \exp\left(u^2\over 2\right) \int_u^\infty
t\exp\left(-{t^2\over 2}\right) H(t,\varphi)dt. \eqno(A.15b)$$

\blankline
\vfill\eject
\baselineskip=10pt
\noindent
{\bf REFERENCES}
\blankline

\nobreak
\refitem {Bean, A. J., Efstathiou, G., Ellis, R. S., Peterson, B. A., \&
Shanks, T. 1983, MNRAS, 205, 605}
\refitem {Bond, J. R., Cole, S., Efstathiou, G., \& Kaiser, N. 1991, ApJ,
379, 440}
\refitem {Bower, R. J. 1991, MNRAS, 248, 332}
\refitem {Bunn, E. F., Scott, D., \& White, M. 1995, ApJ, 441, L9}
\refitem {Carlberg, R. G. 1994, ApJ, 433, 468}
\refitem {Castaing, B., Gagne, Y., \& Hopfinger, E. J. 1990, Physica D,
46, 177}
\refitem {Catelan, P., \& Scherrer, R. J. 1995, ApJ, 445, 1}
\refitem {Cen, R., \& Ostriker, J. P. 1993, ApJ, 417, 415}
\refitem {Cole, S. M., \& Lacey, C. G. 1996, MNRAS, in press}
\refitem {Colless, M., \& Dunn, A. M. 1996, ApJ, 458, 435}
\refitem {Couchman, H. M. P., \& Carlberg, R. G. 1992, ApJ, 389, 453}
\refitem {Crone, M. M., Evrard, A. E., \& Richstone, D. O. 1994, ApJ, 434,
402}
\refitem {Crone, M. M., \& Geller, M. J. 1995, AJ, 110, 21}
\refitem {Davis, M., \& Peebles, P. J. E. 1983, ApJ, 267, 465}
\refitem {Davis, M., Efstathiou, G., Frenk, C. S., \& White, S. D. M. 1985,
ApJ, 292, 371}
\refitem {Diaferio, A., Ramella, M., Geller, M. J., \& Ferrari, A. 1993,
AJ, 105, 2035}
\refitem {Diaferio, A., Geller, M. J., \& Ramella, M. 1994, AJ, 107,
868}
\refitem {Diaferio, A., Geller, M. J., \& Ramella, M. 1995, AJ, 109,
2293}
\refitem {Doe, S. M., Ledlow, M. J., Burns, J. O., \& White, R. A. 1995, AJ,
110, 46}
\refitem {Efstathiou, G. P., Frenk, C. S., White, S. D. M., \& Davis, M. 1988,
MNRAS, 235, 715}
\refitem {Evrard, A. E., Mohr, J. J., Fabricant, D. G., \& Geller, M. J.
1993, ApJ, 419, L9}
\refitem {Evrard, A. E., Summers, F. J., \& Davis, M. 1994, ApJ, 422, 11}
\refitem {Feller, W. 1966, An Introduction to Probability Theory and Its
Application, Vol. II (New York: John Wiley \& Sons)}
\refitem {Fisher, K. B. 1995, ApJ, 448, 494}
\refitem {Fisher, K. B., Davis, M., Strauss, M. A., Yahil, A., \& Huchra,
J. P. 1994a, MNRAS, 266, 50}
\refitem {Fisher, K. B., Davis, M., Strauss, M. A., Yahil, A., \& Huchra,
J. P. 1994b, MNRAS, 267, 927}
\refitem {Frenk, C. S., Evrard, A. E., White, S. D. M., \& Summers, F. J.
1996, ApJ, submitted}
\refitem {Fusco-Femiano, R., \& Menci, N. 1995, ApJ, 449, 431}
\refitem {Gelb, J. M., \& Bertschinger, E. 1994, ApJ, 436, 491}
\refitem {Geller, M. J., \& Peebles, P. J. E. 1973, ApJ, 184, 329}
\refitem {Gunn, J. E., \& Gott, J. R. 1972, ApJ, 176, 1}
\refitem {Hernquist, L. 1987, ApJS, 64, 715}
\refitem {Hernquist, L., Katz, N., \& Weinberg, D. H. 1995, ApJ, 442, 57}
\refitem {Hickson, P. 1993, Ap. Lett. Comm., 29, 1}
\refitem {Ida, S., \& Taguchi, Y.-h. 1996, ApJ, submitted}
\refitem {Jackson, J. D. 1962, Classical Electrodynamics (New York: John
Wiley \& Sons)}
\refitem {Jing, Y. P., \& Fang, L. Z. 1994, ApJ, 432, 438}
\refitem {Jing, Y. P., Mo, H. J., B\"orner, G., \& Fang, L. Z. 1995,
MNRAS, 276, 417}
\refitem {Kauffmann, G., \& White, S. D. M. 1993, MNRAS, 261, 921}
\refitem {King, I.R. 1965, AJ, 70, 376}
\refitem {King, I.R. 1966, AJ, 71, 64}
\refitem {Kofman, L., Bertschinger, E., Gelb, J. M., Nusser, A., \&
Dekel, A. 1994, ApJ, 420, 44}
\refitem {Lacey, C. G., \& Cole, S. M. 1993, MNRAS, 262, 627}
\refitem {Lacey, C. G., \& Cole, S. M. 1994, MNRAS, 271, 676}
\refitem {Lynden-Bell, D. 1967, MNRAS, 136, 101}
\refitem {Mamon, G. A. 1992a, in Distribution of Matter in the Universe,
ed. G. A. Mamon \& D. Gerbal (Observatoire de Paris, Paris), p.
51}
\refitem {Mamon, G. A. 1992b, in Groups of Galaxies, ed. by O. G.
Richter \& K. Borne (ASP: San Francisco), p. 173}
\refitem {Marzke, R. O., Geller, M. J., Da Costa, N. L., \& Huchra, J. P. 1995,
AJ, 110, 477}
\refitem {Mazure, A., et al.
1996, A\&A, submitted}
\refitem {Miesch, M. S., \& Scalo, J. M. 1995, ApJ, 450, L27}
\refitem {Mohr, J. J., Fabricant, D. G., Geller, M. J., Wegner, G.,
Thorstensen, J., \& Richstone, D. O. 1996, ApJ, submitted}
\refitem {Narayan, R., \& White, S. D. M. 1988, MNRAS, 231, 97P}
\refitem {Navarro, J. F., Frenk, C. S., \& White, S. D. M. 1996, ApJ,
in press}
\refitem {Nusser, A., Dekel, A., \& Yahil, A. 1995, ApJ, 449, 439}
\refitem {Padmanabhan, T. 1990, Physics Reports, 188, 285}
\refitem {Peebles, P. J. E. 1976, Ap\&SS, 45, 3}
\refitem {Press, W. H., \& Schechter, P. 1974, ApJ, 187, 425}
\refitem {Ramella, M., Diaferio, A., Geller, M. J., \& Huchra, J. P. 1994,
AJ, 107, 1623}
\refitem {She, Z.-S. 1991, Fluid Dynamics Research, 8, 143}
\refitem {Sheth, R.K. 1996, MNRAS, in press}
\refitem {Shu, F. H. 1978, ApJ, 225, 83}
\refitem {Spitzer, L. 1958, ApJ, 127, 17}
\refitem {van der Marel, R. P., \& Franx, M. 1993, ApJ, 407, 525}
\refitem {Weil, M. L., \& Hernquist, L. 1996, ApJ, in press}
\refitem {Weinberg, M. D. 1994a, AJ, 108, 1398}
\refitem {Weinberg, M. D. 1994b, AJ, 108, 1403}
\refitem {Weinberg, M. D. 1994c, AJ, 108, 1414}
\refitem {West, M. J., Jones, C., \& Forman, W. 1995, ApJ, 451, L5}
\refitem {White, S. D. M. 1993, Les Houches Lectures, in press}
\refitem {White, S. D. M., \& Frenk, C.S. 1991, ApJ, 379, 52}
\refitem {Zabludoff, A. I., Franx, M., \& Geller, M. J. 1993a, ApJ, 419,
47}
\refitem {Zabludoff, A. I., Geller, M. J., Huchra, J. P., \& Ramella, M. 1993b,
AJ, 106, 1301}
\refitem {Zurek, W. H., Quinn, P. J., Salmon, J. K., \& Warren, M. S. 1994,
ApJ, 431, 559}

\end